\documentclass[aip, amsmath,amssymb, reprint, floatfix]{revtex4-1}
\usepackage{dcolumn}
\usepackage{color}
\usepackage{graphicx}
\usepackage{bm}
\usepackage{soul}
\usepackage[utf8]{inputenc}
\usepackage[T1]{fontenc}
\usepackage{mathptmx}
\usepackage{natbib}
\usepackage{multirow}
\usepackage{colortbl}
\usepackage{chngcntr}
\usepackage{upgreek}

\usepackage{amsmath}
\usepackage{amsfonts}
\usepackage{amssymb}

\newcommand{\titre}{Neuromorphic weighted sums with magnetic skyrmions}

\begin{document}

\title{\titre}

\author{Tristan da Câmara Santa Clara Gomes}\thanks{Present address: Institute of Condensed Matter and Nanosciences, Universit\'{e} catholique de Louvain, Place Croix du Sud 1, 1348 Louvain-la-Neuve, Belgium; tristan.dacamara@uclouvain.be}
\affiliation{Laboratoire Albert Fert, CNRS, Thales, Universit\'e Paris-Saclay, 91767 Palaiseau, France}
\author{Yanis Sassi}
\affiliation{Laboratoire Albert Fert, CNRS, Thales, Universit\'e Paris-Saclay, 91767 Palaiseau, France}
\author{D\'{e}dalo Sanz-Hern\'{a}ndez}
\affiliation{Laboratoire Albert Fert, CNRS, Thales, Universit\'e Paris-Saclay, 91767 Palaiseau, France}
\author{Sachin Krishnia}
\affiliation{Laboratoire Albert Fert, CNRS, Thales, Universit\'e Paris-Saclay, 91767 Palaiseau, France}
\author{Sophie Collin}
\affiliation{Laboratoire Albert Fert, CNRS, Thales, Universit\'e Paris-Saclay, 91767 Palaiseau, France}
\author{Marie-Blandine Martin}
\affiliation{Laboratoire Albert Fert, CNRS, Thales, Universit\'e Paris-Saclay, 91767 Palaiseau, France}
\author{Pierre Seneor}
\affiliation{Laboratoire Albert Fert, CNRS, Thales, Universit\'e Paris-Saclay, 91767 Palaiseau, France}
\author{Vincent Cros}\email{vincent.cros@cnrs-thales.fr}
\affiliation{Laboratoire Albert Fert, CNRS, Thales, Universit\'e Paris-Saclay, 91767 Palaiseau, France}
\author{Julie Grollier}\email{julie.grollier@cnrs-thales.fr}
\affiliation{Laboratoire Albert Fert, CNRS, Thales, Universit\'e Paris-Saclay, 91767 Palaiseau, France}
\author{Nicolas Reyren}\email{nicolas.reyren@cnrs-thales.fr}
\affiliation{Laboratoire Albert Fert, CNRS, Thales, Universit\'e Paris-Saclay, 91767 Palaiseau, France}

\begin{abstract}
Integrating magnetic skyrmions into neuromorphic computing could help improve hardware efficiency and computational power. However, developing a scalable implementation of the weighted sum of neuron signals — a core operation in neural networks — has remained a challenge. Here, we show that weighted sum operations can be performed in a compact, biologically-inspired manner by using the non-volatile and particle-like characteristics of magnetic skyrmions that make them easily countable and summable. The skyrmions are electrically generated in numbers proportional to an input with an efficiency given by a non-volatile weight. The chiral particles are then directed using localized current injections to a location where their presence is quantified through non-perturbative electrical measurements. Our experimental demonstration, which currently has two inputs, can be scaled to accommodate multiple inputs and outputs using a crossbar array design, potentially nearing the energy efficiency observed in biological systems.
\end{abstract}

\maketitle

Magnetic skyrmions are topological magnetic solitons behaving like particles. They are stabilized at room temperature in magnetic thin films or heterostructures with optimized magnetic anisotropy and Dzyaloshinskii-Moriya interaction (DMI) \cite{Fert2017, MoreauLuchaire2016, Jiang2015, Woo2016, Boulle2016}. Recent experimental studies have shown that magnetic skyrmions can be nucleated \cite{Jiang2015, Legrand2017, Butner2017, Soumyanarayanan2017, Finizio2019, Wang2020}, moved \cite{Jiang2015, Legrand2017, Woo2016}, annihilated \cite{Woo2018, Yang2021AdvancedMaterials} and electrically detected using the anomalous Hall effect (AHE) \cite{Maccariello2018, Zeissler2018} or tunnelling magnetoresistance \cite{Hanneken2015, Chen2024, Urrestarazu2024}.

Magnetic skyrmions have a range of features that make them suitable for energy-efficient computing operations \cite{Fert2017, Bourianoff2018, Song2020}, such as stability at room temperature, deep sub-micron dimensions, non-volatility, particle-like behaviour and motion at low power. These characteristics align closely with the needs of neuromorphic computing \cite{Grollier2020}, and recent reports have shown that skyrmions can provide various roles in neuromorphic circuits, from acting as artificial synapses \cite{Huang2017, Song2020} and neurons \cite{Sharad2012, Grollier2020, Chen2020} to functioning as stochastic reshufflers \cite{Pinna2018, Zazvorka2019}. They can even facilitate reservoir computing for data classification based on particle dynamics \cite{Bourianoff2018, Prychynenko2018, Raab2022, Yokouchi2023, Sun2023}. However, a fundamental neural network operation — the weighted sum of input neuron signals — is still missing in the context of skyrmions \cite{Grollier2020}.

A weighted sum operation involves multiplying each input, $X_i$ (where $i$ relates to the index), by a tunable factor known as the synaptic weight, $w_i$ (Fig.~\textbf{1}\textbf{a}). The subsequent results, $Y_i$, are then summed to produce the final outcome $Y_\text{tot} = \sum_{i=1}^M Y_i = \sum_{i=1}^M w_i X_i$, where $M$ is the number of inputs. Complementary metal–oxide–semiconductor (CMOS) neuromorphic circuits achieve weighted sums through the combined use of transistors and static random-access memories that have a large silicon footprint (several micrometres squared per synapse), volatile weights, and consume tens of picoJoules per synaptic operation \cite{Frenkel2019}. Biological systems, in contrast, achieve the weighted sum through the nucleation and release of vesicles (black circles in Fig.~\textbf{1}\textbf{b}) under electrical stimulation, followed by the accumulation and detection of the neurotransmitters that they contain (violet points in Fig.~\textbf{1}\textbf{b}) to produce the output spikes. They are extremely efficient with non-volatile synapses of sub-micron dimensions, and an energy cost of approximately 25 fJ per synaptic event corresponding mainly to a vesicle release \cite{Attwell2001, Harris2012}.

\begin{figure*}[ht!]
	\centering
	\includegraphics[scale=1]{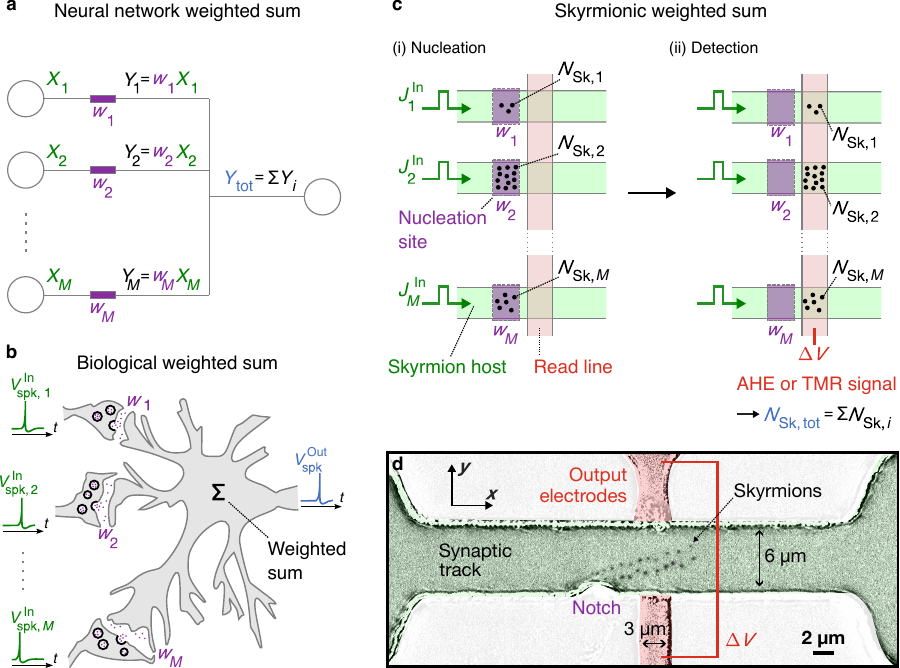}
	\caption{\textbf{Skyrmionic weighted sum: principle and building blocks}. \textbf{a}, Working principle of a	neural network weighted sum that emulates the functionality of the biological neural network. The input values $X_i$ are multiplied by their respective synaptic weights $w_i$ to generate $Y_i$ values that can be summed to produce an output $Y_\text{tot} = \sum_i Y_i$. \textbf{b}, Illustration of a biological weighted summation operation within a neuron. The neuron receives spike voltage signals $V^\text{In}_{\text{spk, }i}$ at various times through synapses, each with its associated synaptic weight $w_i$. This induces the nucleation and motion of vesicles (black circles) containing neurotransmitters (violet dots). These neurotransmitters accumulate at the neuron receptors, resulting in the weighted summation operation $\sum_i w_i V^\text{In}_{\text{spk, }i}$. The neuron generates an output spike $V^\text{Out}_\text{spk}$ when this summation reaches a threshold value, thereby implementing the non-linear activation function. \textbf{c}, Illustration depicting the proposed implementation of a skyrmionic weighted sum, which performs the necessary operation for a neural network weighted sum. Electrical inputs $J^\text{In}_i$ are applied to skyrmion host tracks (in green) to nucleate controlled number of magnetic skyrmions $N_{\text{Sk, } i}$ at nucleation sites, each with an assigned synaptic weight $w_i$ so that $N_{\text{Sk, } i} = w_i J^\text{In}_i$. These skyrmions are moved into a transverse electrical detection zone, established by the intersection of the track and the read line for the electrical output (in red), where the cumulative number of skyrmions $N_\text{Sk, tot} = \sum_i N_{\text{Sk, } i}$ can be electrically detected through an output voltage $\Delta V$ arising from the anomalous Hall effect (AHE) or the tunnel magnetoresistance (TMR). The insets highlight the parallel between the biological weighted sum involving the nucleation and motion of vesicles and the accumulation of neurotransmitters. \textbf{d}, Kerr microscopy difference image of the device with magnetic skyrmions (dark spots), which were nucleated from a notch in a 6-$\upmu$m wide synaptic track made of Ta(5\,nm)/Pt(8\,nm)/[Co(1.2\,nm/Al(3\,nm)/Pt(3\,nm)]$_{10}$ multilayer. Skyrmions moved between the detection electrodes made of 3-$\upmu$m wide and 10-nm thick Ta electrodes connected only at the edges of the track.}
	\label{FigIntro}
\end{figure*}

In this Article, we provide the experimental demonstration of a neuromorphic weighted sum with magnetic skyrmions (Fig.~\textbf{1}\textbf{c}). When an electrical current pulse is injected along a track with index $i$, a corresponding number of skyrmions, $N_{\text{Sk, }i}$, is produced in the nucleation zone. The intention is for this number to be proportional to the input values — signified by the current pulse in each track $J^\text{In}_i$ — multiplied by a factor $w_{i}$, which represents the weights of the weighted sum: $N_{\text{Sk, }i} = w_{i} J^\text{In}_i$. These skyrmions, once generated, can be displaced under the action of electrical pulses of smaller amplitude through the action of spin–orbit torques (SOT) to accumulate in a specific region, enabling a spatial summation expressed as $N_\text{Sk, tot} = \sum_{i} N_{\text{Sk, }i}$. The total number of skyrmions in a selected region can be detected using magnetoresistive means \cite{Maccariello2018, Zeissler2018, Chen2024, Urrestarazu2024} and can therefore implement the weighted sum of inputs:
\begin{equation}
    N_{\text{Sk, }tot} = \sum_{i} w_{i} J^\text{In}_i \quad .
    \label{EqWeightedSum}
\end{equation}
As we demonstrate here, this approach allows for an electrical readout of the total sum. We exploit the unique quasiparticle and non-volatility properties of skyrmions, making them easily countable and summable, and use these properties to perform the weighted sum operation over the number of skyrmions. In this approach, the skyrmions are nucleated and moved in a manner reminiscent of vesicles, and detected and accumulated within neurons in a way analogous to neurotransmitters (inset of Fig.~\textbf{1}\textbf{c}) \cite{Chanaday2019}. Similar functionality could also be achieved with other particle-like magnetic textures like hopfions \cite{Kent2021} or skyrmionics cocoons \cite{Grelier2022}, although their electrical control is currently less advanced.

We show that electrical pulses can nucleate a controllable number of skyrmions from a notch in the host material at room temperature (Fig.~\textbf{1}\textbf{d}). The number of generated skyrmions directly correlates with the count of applied electrical pulses, allowing efficient input encoding (Eq.~1). The weight factor, influenced by the notch’s geometry, can be adjusted by changing the magnetic field applied on the structure (but more practical non-volatile solutions exist). We then non-perturbatively detect the nucleated skyrmions using the AHE. Finally, we showcase an experimental demonstration of the weighted sum with a two-input to one-output configuration. Our skyrmionic weighted sum approach offers non-volatile weights and compactness due to fully-electrical, local operations at a submicron scale. We show that the concept is scalable to multiple inputs and outputs by using a crossbar array-like geometry, with the potential of reaching the energy efficiency found in biological systems.

\section*{Electric pulse controlled nucleation and motion of skyrmions} 

The first experimental challenge to realize the weighted sum is to achieve a precise control of skyrmions nucleation: the quantity of nucleated skyrmions should be directly proportional to the input. This necessitates the development of a reliable method to encode this input within the current pulses that drive the nucleation. Furthermore, the proportionality factor, representing the weight, should be adjustable to enable learning. The second challenge involves ensuring that all nucleated skyrmions are reliably transported from their nucleation site to the designated accumulation and detection region without loss or gain of skyrmions. The incorporation of detection circuits, as depicted by the red area in Figs.~\textbf{1}\textbf{c}-\textbf{d}, can lead to localized alterations in current density, which can impact the motion of skyrmions. Notably, the Hall cross geometry, which is frequently employed for skyrmion detection, can hinder the movement of some or all skyrmions due to reduced current density at its center.

Our building block is shown in Fig.~\textbf{1}\textbf{d} (see Methods for details). The multilayer track is optimized to stabilize isolated skyrmions of approximately 200-nm in diameter at room temperature and to facilitate efficient current-driven skyrmion motion by SOT \cite{krishnia2022}. The skyrmion nucleation site is implemented by a notch that enables a local increase in current density, surpassing the threshold for skyrmion nucleation, while ensuring the rest of the track remains below this threshold \cite{Legrand2017} (Supplementary Information~S1). Skyrmion detection is performed through AHE using a specially crafted Hall cross geometry. We design tantalum (Ta)-based Hall electrodes that are two orders of magnitude more resistive than the skyrmion track, connected solely to its edge to minimize spatial variations of the current density and eliminate potential current leaks that might disrupt skyrmion movement (Supplementary Information S1).

The controlled generation of skyrmions by current injection in our building block device is demonstrated in Fig.~\textbf{2}. As a preparatory step, we orient the magnetisation vertically (along the $+z$ direction) by applying an out-of-plane (OOP) magnetic field of 200\,mT, ensuring complete saturation of magnetisation. Subsequently, by injecting current pulses (current density $J = 160$\,GA/m$^2$ and pulse duration $t = 50$\,ns) under an OOP magnetic field ($\mu_0H_z = 24$\,mT), we selectively nucleate magnetic skyrmions at the notch. Following their formation, these skyrmions are transported, driven by the SOT mechanism \cite{Jiang2015, Legrand2017, Woo2016}, to the designated detection region.

\begin{figure*}[ht!]
	\centering
	\includegraphics[scale=1]{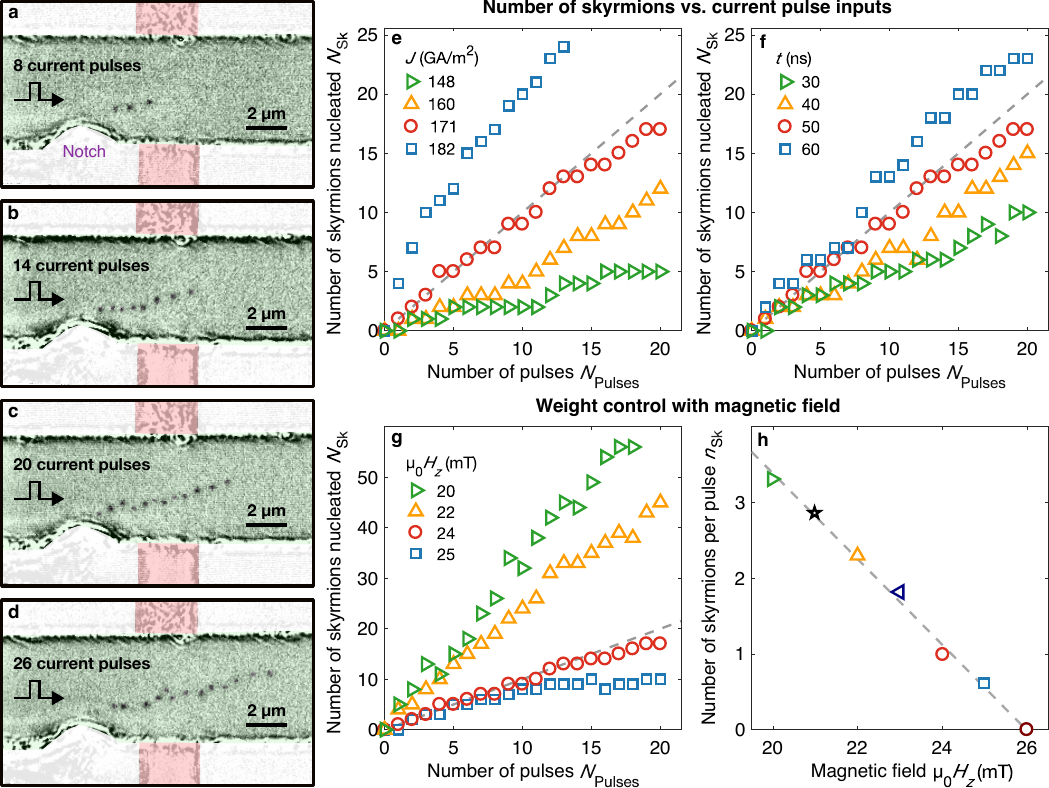}
	\caption{\textbf{Controlled nucleation of magnetic skyrmions at a notch using current pulses as inputs and weight fine-tuning}. \textbf{a}-\textbf{d}, Kerr microscopy images of the device described in Fig.~\textbf{1}\textbf{d}, after the application of 8 (\textbf{a}), 14 (\textbf{b}), 20 (\textbf{c}) and 26 (\textbf{d}) current pulses with $J = 160$\,GA/m$^2$, $t = 50$\,ns, and under an out-of-plane external magnetic field $\mu_0H_z = 24$\,mT. \textbf{e}-\textbf{f}, Number of skyrmions nucleated in the track $N_\text{Sk}$ with the number of pulses applied $N_\text{Pulses}$ (fixed $\mu_0H_z = 24$\,mT) for (\textbf{e}) for several current densities $J$ with fixed $t = 50$\,ns and (\textbf{f}) for several pulse duration $t$ with fixed $J = 171$\,GA/m$^2$. \textbf{g}-\textbf{h}, Control of the weight using the out-of-plane external magnetic fields $H_z$. \textbf{g}, Number of skyrmions nucleated with the number of pulses $N_\text{Pulses}$ for several out-of-plane external magnetic fields $H_z$ with fixed $J = 171$\,GA/m$^2$ and $t = 50$\,ns. \textbf{h}, Linear control of the number of skyrmions nucleated per pulse (slopes in graphs \textbf{e}-\textbf{g}), $n_\text{Sk}$, with the magnetic field $H_z$. The grey dashed lines in \textbf{e}-\textbf{g} correspond to 1 nucleated skyrmion per pulse, in \textbf{h}, it is a linear fit with a slope of $-0.57$\,skyrmion per pulse and per mT. The error bars in (\textbf{h}) have a similar magnitude to that of the symbols.}
	\label{FigControl}
\end{figure*}

The skyrmion count within the track increases linearly with the number of applied current pulses, as demonstrated in Figs.~\textbf{2}\textbf{a}-\textbf{d}. The skyrmions, being nucleated sequentially, form a linear arrangement, following a characteristic oblique trajectory, which is attributed to the inherent topological properties of skyrmions, the skyrmion Hall effect \cite{Jiang2017}. The skyrmion Hall angle, which quantifies the deviation between the current's direction (along $x$) and the skyrmion's propagation path, is approximately 15$^\circ$, consistent with the anticipated angle for this specific material system at the experimental current density \cite{Legrand2017}.

To mitigate any potential adverse effects from such transversal motion, the Hall electrodes are positioned close to the notch. This ensures that skyrmions are detected prior to approaching the track's opposite edge, as visualized in Figs.~\textbf{2}\textbf{a}-\textbf{d}. As expected from the skyrmion Hall angle, we determine that skyrmions typically cover a distance of $\sim$15\,$\upmu$m longitudinally (in the $x$ direction) before nearing the opposing boundary of the track. By analyzing the skyrmion velocity as a function of the position in the track, we observe that the train of skyrmions can completely cross the area of the Ta-based electrodes without any detectable reduction of their velocity and perturbation of their trajectory, because of the much larger sheet resistance of the Ta layer compared to the magnetic multilayer (Supplementary Information S1 and S2). These findings represent a critical advancement in the design of anomalous Hall detection, with significant implications for skyrmion racetrack memories \cite{Fert2013, He2023}, where the perturbation of the skyrmion motion at the detection electrodes can significantly affect the functioning of the device. In the present device, we find a skyrmion velocity ranging from few m/s to few tens of m/s for current density ranging from 150 to 200\,GA/m$^2$. The relation between the number of nucleated skyrmions per pulse and the current density displays a quadratic behaviour, which is a signature of a thermally-driven skyrmion nucleation process (Supplementary Information S3).

\section*{Transforming electrical pulse inputs into skyrmion numbers} 

We demonstrate that our device design enables controlled skyrmion generation, with a linear evolution of their number with the input required to implement the weighted sum operation (Eq.~1). In Figs~\textbf{2}\textbf{e}-\textbf{f}, we display the evolution of the number of nucleated skyrmions $N_\text{Sk}$ in the track observed by Kerr microscopy as a function of the number of current pulses injected in the track $N_\text{Pulses}$, under an OOP external magnetic field $\mu_0H_z = 24$\,mT. 

Our results, presented for varying current pulse densities $J$ in Fig.~\textbf{2}\textbf{e} and different pulse durations $t$ in Fig.~\textbf{2}\textbf{f}, consistently show a direct proportionality between the skyrmion count and the number of applied pulses. This underscores our capability to change the number of skyrmions in the track linearly by adjusting the number of input pulses. The nucleation rate of skyrmions per pulse also displays a linear relationship with pulse duration (Supplementary Information S3, Fig.~\textbf{S4}), offering an alternative to the number of current pulses to encode the input. 

For specific sets of parameters, e.g. $J = 171$\,GA/m$^2$ and $t = 50$\,ns, we observe a one-to-one ratio of skyrmion to pulse (Figs.~\textbf{2}\textbf{e}-\textbf{f}). By modulating these parameters, we can control nucleation probabilities, thus achieving a desired skyrmion number in the track. This flexibility allows for either a small number of skyrmions, facilitating optical detection, or a higher number of skyrmions, suitable for more complex operations. The latter is particularly valuable for addressing device sensitivity issues that may arise from small variations in skyrmion numbers or to reduce potential computation errors caused by unintended skyrmion nucleation or annihilation. A high skyrmion count also improves the electrical signal, making easier the electrical detection. 

\section*{Magnetic field control of the synaptic weight} 

In Fig.~\textbf{2}\textbf{g}, we show that the proportionality factor between the number of nucleated skyrmions $N_\text{Sk}$ and the number of current pulses injected $N_\text{Pulses}$, which in our concept defines the synaptic weight, depends on the amplitude of the external perpendicular magnetic field $H_z$. These results demonstrate that the synaptic weight can be tuned by an external OOP magnetic field. The smaller the value of the external magnetic field $H_z$, the larger is the number of nucleated skyrmions for a given current pulse. Notably, more than 50 skyrmions can be nucleated in the 6-$\upmu$m-wide track in 20 current pulses ($J = 171$\,GA/m$^2$ and $t = 50$\,ns) by decreasing the magnetic field to $20$\,mT. Depending on the application, the number of states could be adjusted by the skyrmion diameter relative to the track width. Here, we focus on accurately assessing device functionality by quantifying the number of skyrmions through Kerr microscopy. To achieve this, 200\,nm large skyrmions are utilized, alongside a limited number that could be confidently counted visually. However, in practical devices, direct skyrmion observation is unnecessary, allowing for the substantial reduction in skyrmion size down in the range of a few tens nanometers in diameter and a corresponding increase in their number.

In Fig.~\textbf{2}\textbf{h}, we present the number of skyrmions nucleated per current pulse, or nucleation probability $n_\text{Sk} = {\rm d}N_\text{Sk}/{\rm d}N_\text{Pulses}$ extracted from linear fits in Fig.~\textbf{2}\textbf{g} as a function of $H_z$. This nucleation probability is found to decrease linearly with $H_z$. A variation of about $-0.57$ skyrmions per pulse and per mT is found up to a maximum field $\mu_0H_{z,\text{max}} = 26$\,mT, where no more skyrmion is nucleated (for fixed $J = 171$\,GA/m$^2$ and $t = 50$\,ns). On the contrary, if the field is further reduced below $20$\,mT, elongated domains are found to be nucleated in the track, thus setting a minimal field for device operation. We estimated the precision on the slope (i.e. the synaptic weights) to be approximately 0.1 skyrmion/pulse (Supplementary Information S4). Therefore, this enables the differentiation of 15 synaptic states for field increments of approximately 0.2 mT in our prototype demonstration. This exceeds the 6 synaptic states achieved with skyrmion in ref.~\onlinecite{Chen2020}, which reached 78\% accuracy in recognizing the MNIST dataset. Notably, they mentioned in their paper that with 13 synaptic states, they anticipate reaching 98.61\% accuracy, which is nearly equivalent to the ideal full-precision synapses. Increasing the number of skyrmions offers a solution to enhance weight precision and expand the attainable number of synaptic states. In this context, reducing skyrmion size allows larger number of particles, making smaller skyrmions preferable to go towards analogue systems.

Our findings indicate that adjusting the magnetic properties at the nucleation site can effectively modulate the synaptic weight. Importantly, it demonstrates the linearity of the synaptic weights control with the magnetic field. In prospective devices, achieving such modifications could be realized in a non-volatile and local way, possibly leveraging magneto-ionic effects, as elaborated below.

\section*{Neuronal detection and output}

\begin{figure*}[ht!]
	\centering
	\includegraphics[scale=1]{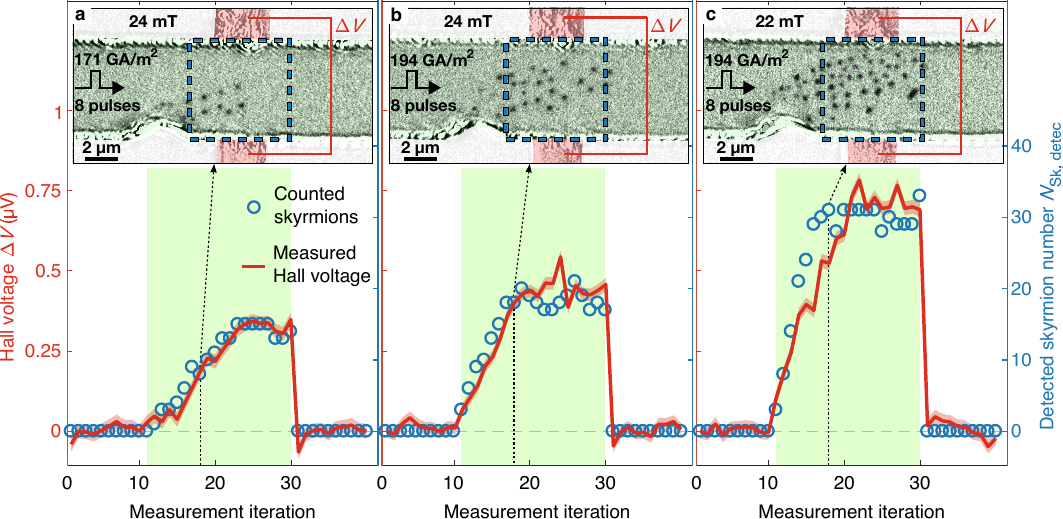}
	\caption{\textbf{Non-perturbative electrical detection of the number of skyrmions using thin Ta electrodes}. \textbf{a}-\textbf{c}, Evolution of the Hall voltage $\Delta V = V - V_\text{sat}$ and the number of detected skyrmions $N_\text{Sk, detec}$ as a function of measurement index (1 pulse applied before each measurements with indices 11 to 30) for pulses with $J=171$\,GA/m$^2$ (\textbf{a}), $194$\,GA/m$^2$ (\textbf{b}-\textbf{c}) (fixed $t = 50$\,ns) and $\mu_0H_z=24$\,mT (\textbf{a}-\textbf{b}) and $22$\,mT (\textbf{c}). Ten measurements of the voltage are performed before the application of the pulses and after magnetically erasing the skyrmions, defining $V_\text{sat}$. The green areas indicate the application of the current pulses. Insets in \textbf{a}-\textbf{c}: corresponding Kerr microscopy images of the nucleated skyrmions after 8 current pulses applied in the different conditions. The blue dashed boxes indicate the detection zone estimated as a 6-$\upmu$m wide square. The shaded red areas indicate the uncertainty of the Hall voltage measurements taken as twice the standard deviation of the initial and final 10 measurements (20 measurements in total).}
	\label{FigDetection}
\end{figure*}

The final summation in our device corresponds in practice to the measurement of the anomalous Hall voltage across the detection area in which skyrmions have been accumulated (Methods and Supplementary Information S2: the Hall voltage hysteresis loop is shown in Figs.~\textbf{S3}\textbf{a}-\textbf{b}). The variation in Hall voltage, denoted as $\Delta V$, is shown in Fig.~\textbf{3} as skyrmions are introduced into the detection zone (blue dotted square in Fig.~\textbf{3}). We use for the visual detection of skyrmions a 6 by 6 $\upmu$m$^2$ box centered on the Hall bar that is assumed to be the electrical detection area (Supplementary Information~S2, Figs.~\textbf{S3}). This choice is motivated by a recent study \cite{FigueiredoPrestes2021}, which demonstrated a consistent anomalous Hall voltage contribution for magnetic textures contained within this specific size, while the influence of magnetic textures positioned outside this box appears to diminish exponentially as the distance from the Hall cross increases. This variation is displayed under three distinct combinations of magnetic fields and currents. The x-axis, titled ``measurement iteration'', represents consecutive measurement sequences (detailed in Methods). After 10 measurements of the saturated voltage value (i.e. $\Delta V \approx 0$), 20 current pulses are used to nucleate skyrmions (green area in Fig.~\textbf{3}). For each set of conditions, $\Delta V$ rises due to the accumulated skyrmions in the detection zone. Then, a saturating field of $200$\,mT is applied to reset the system (eliminates all the skyrmions), followed by 10 final voltage measurements, which revert to $\Delta V \approx 0$. Reversing the current pulse direction is another means (electrical) to erase skyrmions one by one: the skyrmions move back along the same oblique trajectory and get eventually annihilated at the notch (Supplementary Information~S5, Figs.~\textbf{S7}). However, this annihilation mechanism can lead to few residual skyrmions at the notch location, leading to potential small errors in subsequent iterations. The error can be minimized using more pulses, at the expense of a more energetic reset process.

In Fig.~\textbf{3}, each skyrmion contributes equally to reducing the mean magnetization of the initial uniformly magnetized state \cite{Maccariello2018}. This means that the Hall voltage variation, $\Delta V$, can be directly associated with the number of skyrmions present in the detection zone, denoted as $N_\text{Sk, detec}$. Alongside the Hall measurements in Fig.~\textbf{3}, we also plot the number of skyrmions identified in the detection zone via Kerr imaging (illustrated in the insets). We base our analysis on the assumption that the AHE is proportional to the $z$-component of the magnetisation distribution within the 6-$\upmu$m$^2$ square detection area. Skyrmions further away from this center are excluded from consideration. This methodological choice is supported by experimental data showing a proportional relationship between $\Delta V$ values and the count of skyrmions derived from the magneto-optic Kerr images. Notably, the ratio of the scales on the y-axes of Figs.~\textbf{3}\textbf{a}-\textbf{c} are constant across all plots. This showcases the concomitant increase in both skyrmion counts and $\Delta V$ values with either increasing pulse current density (Figs.~\textbf{3}\textbf{a}-\textbf{b}) or decreasing OOP external magnetic field intensity $H_z$ (Figs.~\textbf{3}\textbf{b}-\textbf{c}). Given the limited size of the detection zone, both the skyrmion count and the resultant Hall voltage eventually plateau as the detection area becomes saturated or skyrmions begin to move beyond its boundaries.

From our measurements, we deduce a contribution of approximately $22$\,nV to the Hall voltage variation for each skyrmion (using a dc current of 100\,$\upmu$A) with a standard deviation of 7\,nV (Supplementary Information~S6). This finding is in line with data presented by Maccariello et al. in Ref.~\onlinecite{Maccariello2018}. Additionally, by analyzing the Hall voltage shift during full magnetisation reversal of the detection zone (Supplementary Information~S2, Fig.~\textbf{S3}\textbf{b}) and assuming skyrmions as circular domains with a uniform, opposite "down" magnetisation, we estimate the skyrmion diameter to be approximately $222\pm33$\,nm, as expected for such multilayer track. The estimated standard deviation of the skyrmion diameter primarily arises from uncertainties related to skyrmion counting and electrical noise. Moreover, achieving precise control over the skyrmion diameter is not a fundamental requirement for this study; instead, we focus on harnessing their total number. 

Collectively, these results demonstrate the capability of electrically counting skyrmions with high precision using AHE, a crucial aspect for executing the neuromorphic weighted sum operation. The reproducibility of the skyrmion nucleation and Hall voltage output for fixed parameters is demonstrated in Supplementary Information~S4.

\section*{Demonstration of the weighted sum operation}

\begin{figure*}[ht!]
	\centering
	\includegraphics[scale=1]{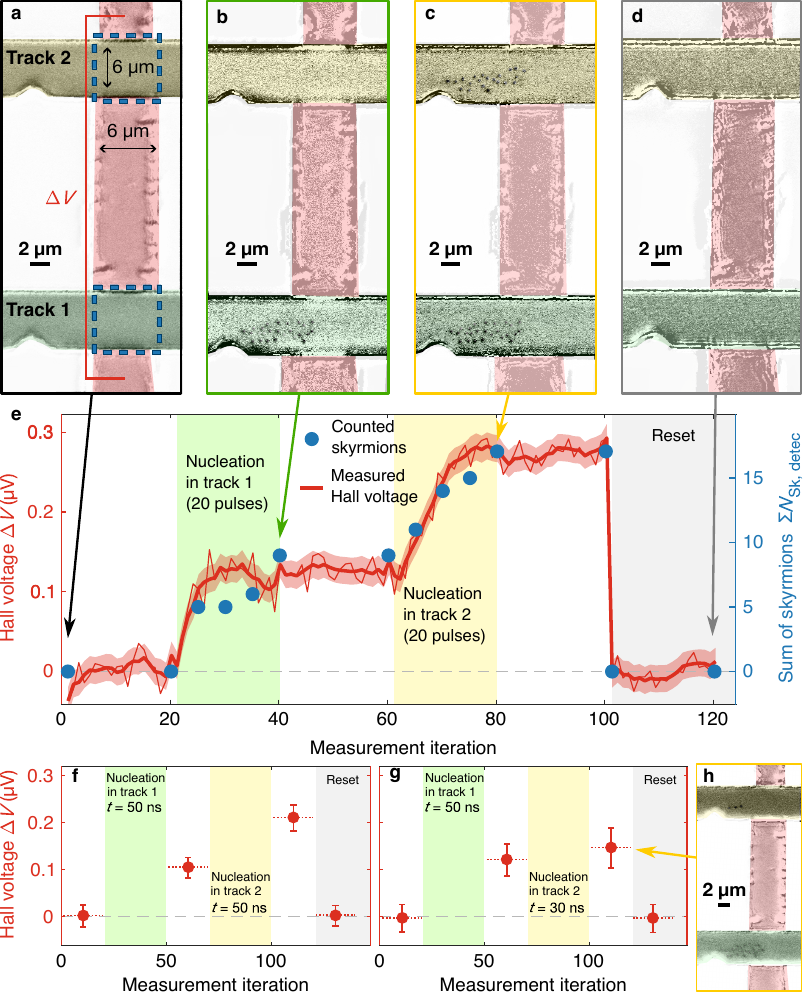}
	\caption{\textbf{Weighted sum in a device composed of two parallel multilayer tracks}. \textbf{a}-\textbf{d}, Kerr microscopy images of the device composed of two 6-$\upmu$m wide parallel magnetic multilayer tracks connected by a transverse 6-$\upmu$m wide Ta Hall electrode. After the saturation of the track using $\mu_0H_z = 200$\,mT (\textbf{a}), skyrmions can be selectively nucleated in track 1 (\textbf{b}) and track 2 (\textbf{c}), before being erased by another $\mu_0H_z = 200$\,mT (\textbf{d}). \textbf{e}, Hall voltage $\Delta V$ (in red) and the corresponding sum of the skyrmion number detected in both tracks $\sum N_\text{Sk, detec}$ (in blue) for successive injection of skyrmion in the tracks. 20 nucleation pulses are successively applied to each track (indicated by the green and yellow areas for track 1 and 2 respectively) using current pulse of about 116\,GA/m$^2$ and 50\,ns, at $\mu_0H_z=20$\,mT. The thin red curve represents the raw electrical measurements after correction of the drift, while the thick one is the same after smoothing. The shaded red areas indicate the uncertainty of the Hall voltage measurements taken as twice the standard deviation of the initial and final 20 measurements (40 measurements in total). \textbf{f}-\textbf{g}, Mean value of the Hall voltage $\Delta V$ after 30 nucleation pulses successively applied to each track, where the pulse duration applied to track 2 is reduced from 50\,ns (\textbf{f}) to 30\,ns (\textbf{g}). \textbf{h}, Kerr microscopy images corresponding to the measurement (\textbf{g}). The error bars in \textbf{f}-\textbf{g} represent the standard deviation of the Hall voltage based on 20 specific measurements, with the range of these measurements indicated by the length of the dotted horizontal line.}
	\label{FigSum}
\end{figure*}

We present the weighted sum operation using a configuration with two inputs and one output within the device shown in Fig.~\textbf{4}\textbf{a}, consisting of two parallel tracks intersected by a Ta Hall electrode (Methods and Supplementary Information~S7). A current pulse generator is connected to the track 1, then to the track 2, to successively inject skyrmions in both tracks (Figs.~\textbf{4}\textbf{b} and \textbf{c}), before reset by applying an external magnetic field $\mu_0H_z=200$\,mT (Fig.~\textbf{4}\textbf{d}). The corresponding Hall voltage variation $\Delta V$ for current pulses of about $116$\,GA/m$^2$ and $50$\,ns with $\mu_0H_z=20$\,mT is shown in Fig.~\textbf{4}\textbf{e}. The measurement sequence is provided in Methods. 20 pulses are applied to track 1, resulting in the concomitant increase of the skyrmions number detected in track 1 with the Hall voltage. The same process is then repeated for track 2, which leads to a further increase of both the total detected skyrmion number and the Hall voltage. The Hall voltage variation measured for each successive skyrmion injection corresponds well to the value expected from individual measurements of the two Hall crosses (Supplementary Information~S7, Figs.~\textbf{S10}\textbf{c}-\textbf{j}). This result demonstrates that the measured Hall voltage $\Delta V$ is, as desired for the final sum operation of our device, directly proportional to the sum of the skyrmion numbers in the two tracks. 

This device produces the same weight for each track as the notches are nominally identical, and the external OOP magnetic field is the same over the entire structure. To study the evolution of the AHE voltage for different weights in each track, we can however modulate the weight - i.e. the number of skyrmions generated by each pulse - by changing the duration of the pulse (Supplementary Information~S7: the linear evolution of the number of skyrmions and the related Hall voltage with pulse duration is shown in Figs.~\textbf{S10}\textbf{c}-\textbf{j}). In Figs.~\textbf{4}\textbf{f}-\textbf{g}, we illustrate this control, keeping constant the number of current pulses (30), the current density $J=116$\,GA/m$^2$, and the external field $\mu_0H_z = 20$\,mT, but varying the pulse duration. In Fig.~\textbf{4}\textbf{f}, the same pulse duration $t=50$\,ns is used in the two tracks, corresponding to identical weights resulting in two consecutive and identical increases in the Hall voltage, with the mean Hall voltage measuring $105\pm$21\,nV and $211\pm$27\,nV following nucleation in track 1 and track 2, respectively. In Fig.~\textbf{4}\textbf{g}, the pulse duration is reduced to $t = 30$\,ns for injecting skyrmions in the track 2, corresponding to reducing the weight close to zero in this track (Supplementary Information~S7, Figs.~\textbf{S10}\textbf{h} and \textbf{j}). As expected, the resulting average Hall voltage after the injection in track 2 ($147\pm$42\,nV) is almost unchanged compared to the value obtained after the nucleation in track 1 ($121\pm$34\,nV). This is equivalent to a suppression of the second term related to track 2 by setting $w_2 \approx 0$, while keeping $w_1$ unchanged, in the total output signal $\Delta V = w_1 N_{\text{Pulse},1} + w_2 N_{\text{Pulse},2}$. The Kerr image reproduced in Fig.~\textbf{4}\textbf{h} explains that in the latter case the detected skyrmion number is close to zero in Track 2 (only side effects).

\section*{Scaling} 

The neuromorphic weighted sum device can be scaled to accommodate more inputs and outputs using a crossbar array geometry, as illustrated in Fig.~\textbf{5}\textbf{a}. By adding more input tracks to a device like the one shown in Fig.~\textbf{4}, the system scales linearly: $M$ parallel tracks can support a weighted sum of $M$ inputs. To increase the number of outputs to $L$, additional notches and Hall bars can be introduced along the track. Utilizing high-resistivity electrodes for AHE measurements, such as the Ta used in our study, can mitigate issues like current leaks and sneak paths that are common in traditional crossbar arrays with oxide materials and conductive electrodes. One advantage of our device, compared to previously proposed neuromorphic devices using skyrmions \cite{Song2020}, is that it requires only 1 transistor per synapse, rather than 2 transistors per synapse, thus reducing the device area and read energy.

To realize a neuromorphic computing device, a non-linear activation function must be applied to the weighted sum results, which has not been yet investigated in the present work. Such function can be applied to the resulting AHE voltage $\Delta V$ by leveraging CMOS technology \cite{Ambrogio2023}, which is compatible with our device, or magnetic tunnel junctions (MTJs) \cite{Torrejon2017, Grollier2020}. In Deep Neural Networks (DNNs), Multiply-and-Accumulate (MAC) operations are the most dominant and thus are crucial to be executed efficiently in hardware \cite{Azghadi2020}.

MTJs could also be used directly for electrical skyrmion detection through their tunnel magnetoresistance (TMR) ratio instead of relying on linear Hall voltages. The device concept is schematically illustrated in Fig.~\textbf{5}\textbf{a}. A large TMR ratio, akin to those in MRAM MTJ bits, could provide a more robust output signal, simplifying on-chip detection circuits. The TMR signal is expected to monotonically increase with the cumulative number of skyrmions beneath the MTJ, and to introduce the necessary non-linearity, especially for large TMR (Supplementary Information~S8, Figs.~\textbf{S11}). Consequently, using MTJs as skyrmion detectors can intrinsically apply the non-linear activation function of the neurons directly at the crossbar outputs. This approach has the potential to execute all essential neuromorphic computing operations within a single device, eliminating the need for external electrical interconnections between the individual neural layer components.

\begin{figure*}[ht!]
	\centering
	\includegraphics[scale=1]{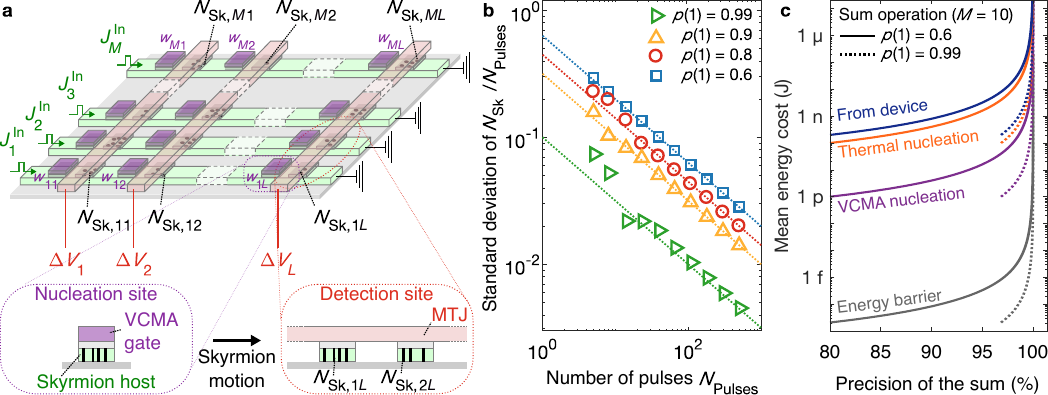}
	\caption{\textbf{Schematic and characteristics of an expanded neuromorphic computing device}. \textbf{a}, Crossbar architecture comprising $M$ parallel tracks (in green) intersected by $L$ Magnetic Tunnel Junctions (MTJ, in red). It enables the introduction of $M$ electrical inputs to inject skyrmions from each nucleation site made of voltage-controlled magnetic anisotropy (VCMA) gate (in purple). The skyrmions are then moved into the detection sites where their number naturally sum, inducing $L$ electrical output voltages. The output voltage produced on column $j$ is given by $\Delta V_{j} = f[\sum_iN_{\text{Sk, }ij}] = f[\sum_i(w_{ij}\,N_{\text{Pulse }i})]$, where $i$ is the track index and $f$ reflects the non-linear activation function of the MTJ applied to the sum of the skyrmion number $\sum_iN_{\text{Sk, }ij}$. \textbf{b}, Evolution of the standard deviation $\sigma$ of the nucleated number of skyrmion $N_\text{Sk}$ over the number input pulses $N_\text{Pulse}$ as a function of $N_\text{Pulse}$ for a single synapse and for several value of the probability of nucleating one skyrmion $p(1)$. The larger the number of attempted skyrmions or the larger $p(1)$, the smaller the relative error with $\sigma \approx \sqrt{ (1-p(1))/N_\text{Pulse}}$. \textbf{c}, Estimated mean energy cost in a device composed of 10 synapses as a function of the desired weighted sum precision considering the energy to nucleate a single skyrmion as $\sim$20\,pJ (estimate from the device in Fig.\,\textbf{2}), 10\,pJ (thermally induced nucleation), 100\,fJ (VCMA induced nucleation) and $500 k_\text{B} T \approx 2$\,aJ at $T = 300$\,K (energy barrier for stable skyrmions).}
	\label{FigFinalDeviceMTJ}
\end{figure*}

In our demonstration, the input is coded by the current pulse number directed to a track, and the Hall voltage serves as the output signal. For future integration of multiple crossbar arrays, consistent types of inputs and outputs would be beneficial. One approach is the use of new spin or orbital torques for skyrmion nucleation \cite{krishnia2023quantifying}, rather than thermal effects. These torques should also result in a linear correlation between the current density and the number of skyrmions. Therefore, the amplitude of the current pulses could act as an input parameter. Conveniently, conversions between such inputs and outputs could be efficiently executed in CMOS using straightforward amplification using transistors. Another promising direction involves utilizing voltage-controlled magnetic anisotropy (VCMA) to induce skyrmion nucleation by applying gate voltage within the nucleation sites \cite{Schott2017, Bhattacharya2020, Urrestarazu2024, Chen2024}. In this scenario, the inputs are directly encoded in voltage amplitudes, ensuring direct consistency with the output voltages.

\section*{Tunable weights} 

Our experimental proof-of-concept device features fixed weights, but we demonstrate that they can be tuned by modifying the local magnetic fields at the nucleation sites. A viable approach for developing miniature trainable devices involves leveraging magneto-ionic effects. By applying electric field gating, these effects can induce local, non-volatile, and reversible alterations to the magnetic anisotropy and DMI within the magnetic multilayer stack \cite{Bernand-Mantel2013, Bauer2015, HerreraDiez2019, Srivastava2018}. Such modifications enable tuning of skyrmion nucleation \cite{Schott2017, Bhattacharya2020} and dynamics \cite{Fillion2022, Dai2023}, effectively adjusting the synaptic weights, in a non-volatile and reversible way \cite{HerreraDiez2019, Mishra2019}. We recently proved the ability to control the magnetic properties and the domain pattern in multi-repeat Pt/Co/Al multilayers similar to those used in this work \cite{Tristan2024}. Moreover, the observed linear correlation between the number of nucleated skyrmions and the OOP external magnetic field, as seen in Fig.~\textbf{2}\textbf{h}, strongly suggests a potential linear relationship between the number of nucleated skyrmions and the sample's anisotropy field. This suggests a straightforward calibration of the synaptic weights. The gate employed for the magneto-ionic modification of magnetic properties can also serve for VCMA-induced skyrmion nucleation with significantly reduced energy requirements \cite{Schott2017, Bhattacharya2020, Urrestarazu2024, Chen2024}. This could be achieved by leveraging different timescales for VCMA nucleation of skyrmions and the In Fig.~\textbf{5}\textbf{a}, we show a schematic representation of the nucleation site design for VCMA-based nucleation and weight control. 

Another approach for implementing non-volatile weights involves harnessing the skyrmion non-volatility to avoid the nucleation step at every computational iteration. Input current pulses would be used exclusively to move skyrmions from the initial nucleation site towards the detection zone, with a number of skyrmions reaching this area that needs to be proportional to the current pulse input. Then, the skyrmions would be moved back to the nucleation site by electric pulses of opposite polarity. By adjusting the total number of skyrmions within the track at the nucleation site, non-volatile weights can be effectively realized. However, a reversible skyrmion motion is required to prevent alteration of the effective weight.

\section*{Energy of synaptic operation} 

The energy for our proposed skyrmionic weighted sum is determined by the nucleation energy of a skyrmion. We experimentally demonstrate this sum operation using a thermal nucleation scheme \cite{Legrand2017}, related to a temperature increase by about $\Delta T \approx 100$\,K, leading to $E_{\rm th} \approx 10$\,pJ/skyrmion (Supplementary Information~S9). Reducing the metallic volume to be heated can further decrease $E_{\rm th}$ to the pJ/skyrmion range. In our experimental device, the heat energy is provided by a current pulse and has been estimated to be approximately 20\,pJ/skyrmion (Supplementary Information~S9), corresponding well to the expected thermal energy required for the nucleation of a skyrmion. In practical devices in which skyrmion counting via Kerr microscopy would be unnecessary, reducing their size decrease the energy consumption and increase compactness with the square of the diameter; for 30-nm diameter skyrmions, we expect a 50-fold reduction. Skyrmions can be scaled down to diameters as small as a few tens of nanometers in magnetic multilayers at room temperature \cite{Legrand2019, Soumyanarayanan2017}, and there is no strict physical limitation to decrease them further in the few-nanometer scale.

Efficient nucleation techniques, such as VCMA, can potentially lower nucleation energy to just under 100-fJ, based on recent MTJ experiments \cite{Chen2024, Urrestarazu2024}. VCMA devices, given their input and weight encoding, promise significant energy savings per synaptic event compared to neuromorphic CMOS chips ($\sim$10\,pJ/synaptic event \cite{Frenkel2019}), with the added benefit of weight retention. Remarkably, this is close to the energy consumption of a biological synaptic event, approximately $\sim$25\,fJ, associated with vesicle release \cite{Attwell2001}, underscoring the potential of skyrmions for brain-equivalent efficiency in neuromorphic computing. Although stochastic thermal nucleation is used in this work, deterministic generation is suitable as the only requirement is the linearity. VCMA nucleation could avoid the requirement of skyrmion motion between nucleation and detection sites by nucleating directly the skyrmions through voltage within a MTJ \cite{Urrestarazu2024, Chen2024}. This nucleation method also mitigates the adverse effects of thermal fluctuations in our system, which might otherwise affect the motion of skyrmions, including the skyrmion Hall angle \cite{Jiang2017, Juge2019, Litzius2020}. Additionally, skyrmions can also be annihilated by gate voltage in MTJ, also offering an electrical reset method when MTJ are employed \cite{Chen2024}. Finally, the lowest energy for skyrmion nucleation that can be expected is approximately $500 k_\text{B} T \approx 2$\,aJ at $T = 300$\,K, corresponding to the energy barrier to stabilize skyrmions.

In addition to the skyrmion nucleation energy contribution, the comprehensive evaluation of energy consumption in the device necessitates consideration of supplementary factors, including the energy expended during the electrical read-out and reset of the skyrmion number and the operation of the CMOS circuit. However, the exhaustive estimation and optimization of these energy components lie beyond the scope of this study. The read-out energy may be reduced through the utilization of optimized MTJs with substantial TMR ratio, coupled with operation at high frequencies, which could lead to value below the fJ \cite{He2017}. It should also be emphasized that these improvements will be highly challenging to achieve, with several cross-compatibility issues and necessary compromises between the proposed improvement solutions. For instance, combining nucleation and detection sites at the same location requires development of gate materials that simultaneously provide a large TMR signal, non-volatile magnetic property changes via magneto-ionic gating, and VCMA skyrmion nucleation, presenting a material challenge due to the competing requirements of these functionalities. In the case of separated nucleation and detection sites, development is required to facilitate the displacement of skyrmions from one site to the other. Notably, the processes of nucleation at one MTJ, displacement, and detection at a separate MTJ have already been demonstrated for domain walls \cite{Raymenants2021}.

\section*{Reliability of the sum operation} 

The thermal origin of skyrmion nucleation makes the process stochastic. If the system is tuned to produce 1 skyrmion per electrical pulse, the outcome can be different than one with a probability $p(\bar{1})$. In our experimental devices, we observe the nucleation of zero, one or two skyrmions when the current pulses are shaped to create a single skyrmion/pulse in average. We have performed numerical simulations for several values of $p(1)$ to determine the relative deviation $\sigma$ of the number of nucleated skyrmion $N_\text{Sk}$ for $N_\text{Pulse}$ electrical pulses (Supplementary Information~S10). Assuming that $p(0)=p(2)=p(\bar{1})/2$, the results show an excellent agreement with the formula $\sigma \approx \sqrt{p(\bar{1})/N_\text{Pulse}}$ (Fig.~\textbf{5}\textbf{b}). In our prototype, we estimate $p(1)$ to be approximately 0.6, a number which also accounts for the detection noise (Supplementary Information~S10 for all details). This number is comparable to the probability of vesicle release in biology (corresponding to the skyrmion nucleation in our concept) typically between 0.25 to 0.5 \cite{Harris2012}. 

The precision of the sum operation is set by this nucleation probability. When $M$ independent synapses produce each one $N_{\text{Sk, } m}$, the expected standard deviation of the sum, $\sqrt{\Sigma_{m=1}^{M}\sigma^2(m)}$, is equal to $\sqrt{M}\sigma$ if all synapses have the same standard deviation. When $N_{\rm Pulse}$ are injected per synapse, the sum is equal to $M\,N_{\rm Pulse}$, the precision of the sum is given by $1-\sigma(N_{\rm Pulse})/\sqrt{M} = 1-\sqrt{p(\bar{1})/(M N_{\rm Pulse})}$ and the mean energy cost is $M\,N_{\rm Pulse}$ times the estimated skyrmion nucleation energy. In Fig.~\textbf{5}\textbf{c}, we depict the relationship between the sum precision and the mean energy cost for 10 synaptic operations ($M =$ 10). Achieving higher precision necessitates greater energy consumption. Consequently, an optimal operating point can be identified by balancing energy cost against sum precision requirements akin to a trade-off encountered in biology \cite{Harris2012}. As seen in Fig.~\textbf{5}\textbf{c}, relatively high precision on the sum operation can be achieved at low energy cost, even using $p(1) \approx 0.6$. Additionally, deep neural networks architectures can be robust against local inaccuracies, which can be absorbed by the network. Two alternative paths for device optimization are (1) to encode numbers with larger number of skyrmions, or (2) to improve the deterministic nucleation of skyrmions. Other nucleation schemes, for example inspired by MRAM technologies, might reduce $p(\bar{1})$ by many orders of magnitude. For 10 synaptic operations, the lowest physical limit for the energy required for skyrmion nucleation is in the range of 10 aJ to 1 fJ, and a single synaptic operation is in the 1-100 aJ range.

\section*{Conclusion} 

We have reported a neuromorphic weighted sum based on magnetic skyrmions. We first developed a method for electrically controlling skyrmion nucleation and movement within specially crafted magnetic multilayered tracks. The linear association between the number of generated skyrmions and applied current pulses, under various sets of parameters, implements the multiplication of the inputs by a synaptic weight. The weight can be finely tuned through minor adjustments to the out-of-plane external magnetic field (close to $-0.57$ skyrmions per pulse and per mT). Using optimized parameters, we demonstrated the injection of over 50 skyrmions into a 6-$\upmu$m-wide track.

Our design employs highly resistive tantalum-based transverse electrodes connected to the edge of the track only, enabling the electrical detection of skyrmions via the anomalous Hall effect without adversely affecting their motion or providing sneak paths in crossbar arrays. We validated the weighted sum operation in a device featuring two parallel tracks intersected by a Hall electrode; the resulting Hall voltage corresponds to the combined number of skyrmions in both tracks. This ensures efficient execution of the fundamental weighted sum operation, a cornerstone for neuromorphic computing.

Looking forward, integrating magneto-ionic effects for non-volatile, reversible control over local magnetic properties, along with non-linear electrical detection via magnetic tunnel junctions, could provide comprehensive functionality for neuromorphic computing. The approach could, ultimately, lead to brain-inspired hardware capable of weighted sums at an energy expenditure as minimal as 1-100\,aJ per synaptic operation.

\section*{Methods}
\subsection*{Device fabrication}

The multilayered stack used to stabilize skyrmions at room temperature: Ta(5\,nm)/Pt(8\,nm)/[Co(1.2\,nm/Al(3\,nm) /Pt(3\,nm)]$_\text{10}$, with the index '10' signifying ten repetitions of the trilayer, is grown by magnetron sputtering at room temperature on Si$_3$N$_4$ substrates. This multilayer stack leverages both interfacial DMI and perpendicular magnetic anisotropy (PMA) allowing to stabilize skyrmions of approximately 200-nm in diameter at room temperature. The Si$_3$N$_4$ substrates are used to enhance thermal energy dissipation and minimize the number of nucleated skyrmions in our demonstrative device, allowing for an easy visual counting and enhancing visual observation of the device operation. The device geometries are defined by a three steps UV lithography process. A first step is used to define the track geometry (in green in Figs.~\textbf{1}\textbf{d}), as well as contacts for the Hall electrodes, which is obtained by Ar$+$ ion etching of the multilayer film. The track shown in green in the Kerr microscopy images of Figs.~\textbf{2}\textbf{a}-\textbf{d}, spans 6\,$\upmu$m in width and 40\,$\upmu$m in length, bearing an impedance of 98\,$\Omega$ including the contacts. It incorporates a rounded triangular nucleation notch, accounting for roughly 17\% of the track’s width. A second step is used to define the crossing Hall bar electrodes (in red in Figs.~ \textbf{1}\textbf{d}), still by UV lithography, while keeping the resist mask used for the first step to ensure that the electrodes are only connected on the edge of the multilayer track. The 10-nm thick Ta electrodes are then deposited by sputtering and lifted-off, measuring 3\,$\upmu$m in width for the devices in Figs.~\textbf{2} and~\textbf{3}. Finally, a third lithography step is used to lift-off sputtered Ti (10\,nm)/Au (200\,nm) electrical contacts for electrical measurements. The proof-of-concept device used to demonstrate the weighted sum operation, shown in Fig.~\textbf{4}\textbf{a}, consists of two parallel magnetic multilayered tracks, each 6-$\upmu$m wide and equipped with a nucleation notch. These tracks are intersected by a 6-$\upmu$m wide Ta Hall electrode.

\subsection*{Kerr microscopy}

Kerr microscopy is used to observe and count the skyrmions in the devices. For each measurement series, a reference image is taken in the saturation state and subtracted to the displayed image. This image difference enhances the magnetic contrast, and reduces the "topographical" features. The light source is a blue light emitting diode (LED) with a peak intensity at $445$\,nm wavelength. For the maximum resolution, we use a 100x lens in immersion in a high refraction index oil ($n=1.5$) and large numerical aperture ($1.3$). Due to thermal drifts, the sample is realigned before each image using a 3D piezoelectric stage. The focus is a key parameter to get the contrast of the sub-wavelength diameter skyrmions. A fine focus series of images is taken varying the focus in order to achieve maximum resolution. The best-focused image is automatically selected, and its optimized focus position is then forwarded for the next image acquisition. Due to this procedure, each acquisition takes typically a few minutes. A final fine correction for the drift in planar directions is made in post-treatment.

The color of the Kerr images presented in this work have been modified to improve the readability of the figures, the Kerr images without color modifications are provided in the Supplementary Information~S11.

\subsection*{Electrical measurements}

Skyrmions are nucleated and moved using pulses generated by an arbitrary waveform generator connected to a $38$\,dB amplifier, typical voltages are in the volt range. The Hall voltage measurements are performed by using a dc current source and a nanovoltmeter to measure the transverse Hall voltage with 100\,$\upmu$A. This current corresponds to a current density in the range of $0.1$\,GA/m$^2$, which is about three orders of magnitude below the threshold current for skyrmion nucleation and motion. To improve the electrical measurement precision, each voltage measurement is the average of 50 individual pair of measurements with positive and negative 100\,$\upmu$A current with duration of $100$\,ms. The measured transverse voltage is considered as proportional to the mean magnetisation along the $z$ direction \cite{Maccariello2018}. AHE loop measurements are obtained by sweeping an OOP external magnetic field and recording the Hall voltage to which an offset is subtracted. The offset corresponds to a small longitudinal voltage component (Supplementary Information~S2, Fig.~\textbf{S3}).

The measurement sequence for the building block device in Fig.~\textbf{3} is the following: Initially, 10 voltage measurements are taken without any skyrmion injections to determine the saturated magnetization value, meaning when no skyrmions are present. In the depicted graph, these initial 10 data points exhibit a consistent Hall voltage, with $\Delta V \approx 0$. Following this, 20 current pulses are used to nucleate skyrmions. After each pulse, both the Hall voltage and a Kerr image are recorded. To conclude the measurement sequence, a saturating field of $200$\,mT is applied and 10 voltage measurements are again performed at saturation to correct for voltage drifts.

For the measurements of a single device composed of parallel multilayer tracks, the tracks are connected in parallel to a dc current source. Track 1 and 2 have respectively a resistance value of 130\,$\Omega$ and 120\,$\Omega$. Each track has $12$\,k$\Omega$ calibrated resistances in series at each end of each track. This ensures identical (within 0.1\%) dc current of 100\,$\upmu$A flowing in each track. Moreover, it guarantees that the voltage measured at the edge of the Hall cross only arises from the device instead of the electrical circuit used for the measurement (see Supplementary Information~S7, Fig.~\textbf{S10}\textbf{a} for a schematics of the electrical circuit). With this circuit, the Hall voltage measurement is found to be, indeed, the sum of the individual Hall voltage of each individual track (Supplementary Information~S7, Fig.~\textbf{S10}\textbf{b}). This indicates that the Hall voltage measured at the edge of the device accounts for the sum of the magnetisation contribution of the two Hall crosses. Bias Tees are used to selectively inject current pulses or dc current within the desired track. The measurement sequence for the prototype device in Fig.~\textbf{4} is the same as for the individual track i.e. 20 measurements of the Hall voltage are taken after fully saturating the tracks' magnetisation and a Kerr image is taken as reference. Then, 20 pulses are applied to track 1, each pulse being followed by a voltage measurement, while Kerr images are taken after every 5 pulses. 20 more voltage measurements are carried out to check the voltage stability. The same process is then repeated for track 2. Finally, a saturating field of $200$\,mT is applied and 10 voltage measurements are taken.

The quoted current densities are the averaged ones estimated by dividing the total current by the width of the track (6\,$\upmu$m) and the total magnetic multilayer thickness (85\,nm).

~\\
\noindent
\textbf{Data Availability}\\
The data that support the findings of this study are available at https://doi.org/10.5281/zenodo.13988409. Other relevant data are available from the corresponding authors.\\

\noindent
\textbf{Acknowledgements}\\
This work gets supports from the Horizon2020 Framework Program of the European Commission under FET-Proactive Grant SKYTOP (824123), by the European Research Council advanced grant GrenaDyn (reference: 101020684), by the EU project SkyANN (reference : 101135729) and from a France 2030 government grant managed by the French National Research Agency (ANR-22-EXSP-0002 PEPR SPIN CHIREX).\\

\noindent
\textbf{Authors contribution Statement}\\
M.-B.M., P.S., V.C., J.G. and N.R. conceived the project. N.R., D.S.-H., Y.S and T.d.C.S.C.G. design the measurement procedure. S.C, Y.S and T.d.C.S.C.G. grew the multilayer films and Ta electrodes. T.d.C.S.C.G. patterned the samples, acquired the MOKE and transport data, treated and analysed the data with support from Y.S, D.S.-H., S.K, M.-B.M., P.S., V.C., J.G. and N.R. T.d.C.S.C.G., V.C., J.G. and N.R. prepared the manuscript. All authors discussed and commented the manuscript.\\

\noindent
\textbf{Competing Interests Statement}\\
The authors declare no competing interests.\\

\section*{Reference}


\begin{thebibliography}{59}%
	\makeatletter
	\providecommand \@ifxundefined [1]{%
		\@ifx{#1\undefined}
	}%
	\providecommand \@ifnum [1]{%
		\ifnum #1\expandafter \@firstoftwo
		\else \expandafter \@secondoftwo
		\fi
	}%
	\providecommand \@ifx [1]{%
		\ifx #1\expandafter \@firstoftwo
		\else \expandafter \@secondoftwo
		\fi
	}%
	\providecommand \natexlab [1]{#1}%
	\providecommand \enquote  [1]{``#1''}%
	\providecommand \bibnamefont  [1]{#1}%
	\providecommand \bibfnamefont [1]{#1}%
	\providecommand \citenamefont [1]{#1}%
	\providecommand \href@noop [0]{\@secondoftwo}%
	\providecommand \href [0]{\begingroup \@sanitize@url \@href}%
	\providecommand \@href[1]{\@@startlink{#1}\@@href}%
	\providecommand \@@href[1]{\endgroup#1\@@endlink}%
	\providecommand \@sanitize@url [0]{\catcode `\\12\catcode `\$12\catcode
		`\&12\catcode `\#12\catcode `\^12\catcode `\_12\catcode `\%12\relax}%
	\providecommand \@@startlink[1]{}%
	\providecommand \@@endlink[0]{}%
	\providecommand \url  [0]{\begingroup\@sanitize@url \@url }%
	\providecommand \@url [1]{\endgroup\@href {#1}{\urlprefix }}%
	\providecommand \urlprefix  [0]{URL }%
	\providecommand \Eprint [0]{\href }%
	\providecommand \doibase [0]{http://dx.doi.org/}%
	\providecommand \selectlanguage [0]{\@gobble}%
	\providecommand \bibinfo  [0]{\@secondoftwo}%
	\providecommand \bibfield  [0]{\@secondoftwo}%
	\providecommand \translation [1]{[#1]}%
	\providecommand \BibitemOpen [0]{}%
	\providecommand \bibitemStop [0]{}%
	\providecommand \bibitemNoStop [0]{.\EOS\space}%
	\providecommand \EOS [0]{\spacefactor3000\relax}%
	\providecommand \BibitemShut  [1]{\csname bibitem#1\endcsname}%
	\let\auto@bib@innerbib\@empty
	\bibitem [{\citenamefont {Fert}, \citenamefont {Reyren},\ and\ \citenamefont
		{Cros}(2017)}]{Fert2017}%
	\BibitemOpen
	\bibfield  {author} {\bibinfo {author} {\bibfnamefont {A.}~\bibnamefont
			{Fert}}, \bibinfo {author} {\bibfnamefont {N.}~\bibnamefont {Reyren}}, \ and\
		\bibinfo {author} {\bibfnamefont {V.}~\bibnamefont {Cros}},\ }\bibfield
	{title} {\enquote {\bibinfo {title} {Magnetic skyrmions: advances in physics
				and potential applications},}\ }\href {\doibase 10.1038/natrevmats.2017.31}
	{\bibfield  {journal} {\bibinfo  {journal} {Nature Reviews Materials}\
		}\textbf {\bibinfo {volume} {2}},\ \bibinfo {pages} {17031} (\bibinfo {year}
		{2017})}\BibitemShut {NoStop}%
	\bibitem [{\citenamefont {Moreau-Luchaire}\ \emph {et~al.}(2016)\citenamefont
		{Moreau-Luchaire}, \citenamefont {Moutafis}, \citenamefont {Reyren},
		\citenamefont {Sampaio}, \citenamefont {Vaz}, \citenamefont {Van~Horne},
		\citenamefont {Bouzehouane}, \citenamefont {Garcia}, \citenamefont
		{Deranlot}, \citenamefont {Warnicke}, \citenamefont {Wohlh{\"u}ter},
		\citenamefont {George}, \citenamefont {Weigand}, \citenamefont {Raabe},
		\citenamefont {Cros},\ and\ \citenamefont {Fert}}]{MoreauLuchaire2016}%
	\BibitemOpen
	\bibfield  {author} {\bibinfo {author} {\bibfnamefont {C.}~\bibnamefont
			{Moreau-Luchaire}}, \bibinfo {author} {\bibfnamefont {C.}~\bibnamefont
			{Moutafis}}, \bibinfo {author} {\bibfnamefont {N.}~\bibnamefont {Reyren}},
		\bibinfo {author} {\bibfnamefont {J.}~\bibnamefont {Sampaio}}, \bibinfo
		{author} {\bibfnamefont {C.~A.~F.}\ \bibnamefont {Vaz}}, \bibinfo {author}
		{\bibfnamefont {N.}~\bibnamefont {Van~Horne}}, \bibinfo {author}
		{\bibfnamefont {K.}~\bibnamefont {Bouzehouane}}, \bibinfo {author}
		{\bibfnamefont {K.}~\bibnamefont {Garcia}}, \bibinfo {author} {\bibfnamefont
			{C.}~\bibnamefont {Deranlot}}, \bibinfo {author} {\bibfnamefont
			{P.}~\bibnamefont {Warnicke}}, \bibinfo {author} {\bibfnamefont
			{P.}~\bibnamefont {Wohlh{\"u}ter}}, \bibinfo {author} {\bibfnamefont {J.~M.}\
			\bibnamefont {George}}, \bibinfo {author} {\bibfnamefont {M.}~\bibnamefont
			{Weigand}}, \bibinfo {author} {\bibfnamefont {J.}~\bibnamefont {Raabe}},
		\bibinfo {author} {\bibfnamefont {V.}~\bibnamefont {Cros}}, \ and\ \bibinfo
		{author} {\bibfnamefont {A.}~\bibnamefont {Fert}},\ }\bibfield  {title}
	{\enquote {\bibinfo {title} {Additive interfacial chiral interaction in
				multilayers for stabilization of small individual skyrmions at room
				temperature},}\ }\href {\doibase 10.1038/nnano.2015.313} {\bibfield
		{journal} {\bibinfo  {journal} {Nature Nanotechnology}\ }\textbf {\bibinfo
			{volume} {11}},\ \bibinfo {pages} {444--448} (\bibinfo {year}
		{2016})}\BibitemShut {NoStop}%
	\bibitem [{\citenamefont {Jiang}\ \emph {et~al.}(2015)\citenamefont {Jiang},
		\citenamefont {Upadhyaya}, \citenamefont {Zhang}, \citenamefont {Yu},
		\citenamefont {Jungfleisch}, \citenamefont {Fradin}, \citenamefont {Pearson},
		\citenamefont {Tserkovnyak}, \citenamefont {Wang}, \citenamefont {Heinonen},
		\citenamefont {te~Velthuis},\ and\ \citenamefont {Hoffmann}}]{Jiang2015}%
	\BibitemOpen
	\bibfield  {author} {\bibinfo {author} {\bibfnamefont {W.}~\bibnamefont
			{Jiang}}, \bibinfo {author} {\bibfnamefont {P.}~\bibnamefont {Upadhyaya}},
		\bibinfo {author} {\bibfnamefont {W.}~\bibnamefont {Zhang}}, \bibinfo
		{author} {\bibfnamefont {G.}~\bibnamefont {Yu}}, \bibinfo {author}
		{\bibfnamefont {M.~B.}\ \bibnamefont {Jungfleisch}}, \bibinfo {author}
		{\bibfnamefont {F.~Y.}\ \bibnamefont {Fradin}}, \bibinfo {author}
		{\bibfnamefont {J.~E.}\ \bibnamefont {Pearson}}, \bibinfo {author}
		{\bibfnamefont {Y.}~\bibnamefont {Tserkovnyak}}, \bibinfo {author}
		{\bibfnamefont {K.~L.}\ \bibnamefont {Wang}}, \bibinfo {author}
		{\bibfnamefont {O.}~\bibnamefont {Heinonen}}, \bibinfo {author}
		{\bibfnamefont {S.~G.~E.}\ \bibnamefont {te~Velthuis}}, \ and\ \bibinfo
		{author} {\bibfnamefont {A.}~\bibnamefont {Hoffmann}},\ }\bibfield  {title}
	{\enquote {\bibinfo {title} {Blowing magnetic skyrmion bubbles},}\ }\href
	{\doibase 10.1126/science.aaa1442} {\bibfield  {journal} {\bibinfo  {journal}
			{Science}\ }\textbf {\bibinfo {volume} {349}},\ \bibinfo {pages} {283--286}
		(\bibinfo {year} {2015})}\BibitemShut {NoStop}%
	\bibitem [{\citenamefont {Woo}\ \emph {et~al.}(2016)\citenamefont {Woo},
		\citenamefont {Litzius}, \citenamefont {Kr{\"u}ger}, \citenamefont {Im},
		\citenamefont {Caretta}, \citenamefont {Richter}, \citenamefont {Mann},
		\citenamefont {Krone}, \citenamefont {Reeve}, \citenamefont {Weigand},
		\citenamefont {Agrawal}, \citenamefont {Lemesh}, \citenamefont {Mawass},
		\citenamefont {Fischer}, \citenamefont {Kl{\"a}ui},\ and\ \citenamefont
		{Beach}}]{Woo2016}%
	\BibitemOpen
	\bibfield  {author} {\bibinfo {author} {\bibfnamefont {S.}~\bibnamefont
			{Woo}}, \bibinfo {author} {\bibfnamefont {K.}~\bibnamefont {Litzius}},
		\bibinfo {author} {\bibfnamefont {B.}~\bibnamefont {Kr{\"u}ger}}, \bibinfo
		{author} {\bibfnamefont {M.-Y.}\ \bibnamefont {Im}}, \bibinfo {author}
		{\bibfnamefont {L.}~\bibnamefont {Caretta}}, \bibinfo {author} {\bibfnamefont
			{K.}~\bibnamefont {Richter}}, \bibinfo {author} {\bibfnamefont
			{M.}~\bibnamefont {Mann}}, \bibinfo {author} {\bibfnamefont {A.}~\bibnamefont
			{Krone}}, \bibinfo {author} {\bibfnamefont {R.~M.}\ \bibnamefont {Reeve}},
		\bibinfo {author} {\bibfnamefont {M.}~\bibnamefont {Weigand}}, \bibinfo
		{author} {\bibfnamefont {P.}~\bibnamefont {Agrawal}}, \bibinfo {author}
		{\bibfnamefont {I.}~\bibnamefont {Lemesh}}, \bibinfo {author} {\bibfnamefont
			{M.-A.}\ \bibnamefont {Mawass}}, \bibinfo {author} {\bibfnamefont
			{P.}~\bibnamefont {Fischer}}, \bibinfo {author} {\bibfnamefont
			{M.}~\bibnamefont {Kl{\"a}ui}}, \ and\ \bibinfo {author} {\bibfnamefont
			{G.~S.~D.}\ \bibnamefont {Beach}},\ }\bibfield  {title} {\enquote {\bibinfo
			{title} {Observation of room-temperature magnetic skyrmions and their
				current-driven dynamics in ultrathin metallic ferromagnets},}\ }\href
	{\doibase 10.1038/nmat4593} {\bibfield  {journal} {\bibinfo  {journal}
			{Nature Materials}\ }\textbf {\bibinfo {volume} {15}},\ \bibinfo {pages}
		{501--506} (\bibinfo {year} {2016})}\BibitemShut {NoStop}%
	\bibitem [{\citenamefont {Boulle}\ \emph {et~al.}(2016)\citenamefont {Boulle},
		\citenamefont {Vogel}, \citenamefont {Yang}, \citenamefont {Pizzini},
		\citenamefont {de~Souza~Chaves}, \citenamefont {Locatelli}, \citenamefont
		{Mente{\c s}}, \citenamefont {Sala}, \citenamefont {Buda-Prejbeanu},
		\citenamefont {Klein}, \citenamefont {Belmeguenai}, \citenamefont
		{Roussign{\'e}}, \citenamefont {Stashkevich}, \citenamefont {Ch{\'e}rif},
		\citenamefont {Aballe}, \citenamefont {Foerster}, \citenamefont {Chshiev},
		\citenamefont {Auffret}, \citenamefont {Miron},\ and\ \citenamefont
		{Gaudin}}]{Boulle2016}%
	\BibitemOpen
	\bibfield  {author} {\bibinfo {author} {\bibfnamefont {O.}~\bibnamefont
			{Boulle}}, \bibinfo {author} {\bibfnamefont {J.}~\bibnamefont {Vogel}},
		\bibinfo {author} {\bibfnamefont {H.}~\bibnamefont {Yang}}, \bibinfo {author}
		{\bibfnamefont {S.}~\bibnamefont {Pizzini}}, \bibinfo {author} {\bibfnamefont
			{D.}~\bibnamefont {de~Souza~Chaves}}, \bibinfo {author} {\bibfnamefont
			{A.}~\bibnamefont {Locatelli}}, \bibinfo {author} {\bibfnamefont {T.~O.}\
			\bibnamefont {Mente{\c s}}}, \bibinfo {author} {\bibfnamefont
			{A.}~\bibnamefont {Sala}}, \bibinfo {author} {\bibfnamefont {L.~D.}\
			\bibnamefont {Buda-Prejbeanu}}, \bibinfo {author} {\bibfnamefont
			{O.}~\bibnamefont {Klein}}, \bibinfo {author} {\bibfnamefont
			{M.}~\bibnamefont {Belmeguenai}}, \bibinfo {author} {\bibfnamefont
			{Y.}~\bibnamefont {Roussign{\'e}}}, \bibinfo {author} {\bibfnamefont
			{A.}~\bibnamefont {Stashkevich}}, \bibinfo {author} {\bibfnamefont {S.~M.}\
			\bibnamefont {Ch{\'e}rif}}, \bibinfo {author} {\bibfnamefont
			{L.}~\bibnamefont {Aballe}}, \bibinfo {author} {\bibfnamefont
			{M.}~\bibnamefont {Foerster}}, \bibinfo {author} {\bibfnamefont
			{M.}~\bibnamefont {Chshiev}}, \bibinfo {author} {\bibfnamefont
			{S.}~\bibnamefont {Auffret}}, \bibinfo {author} {\bibfnamefont {I.~M.}\
			\bibnamefont {Miron}}, \ and\ \bibinfo {author} {\bibfnamefont
			{G.}~\bibnamefont {Gaudin}},\ }\bibfield  {title} {\enquote {\bibinfo {title}
			{Room-temperature chiral magnetic skyrmions in ultrathin magnetic
				nanostructures},}\ }\href {\doibase 10.1038/nnano.2015.315} {\bibfield
		{journal} {\bibinfo  {journal} {Nature Nanotechnology}\ }\textbf {\bibinfo
			{volume} {11}},\ \bibinfo {pages} {449--454} (\bibinfo {year}
		{2016})}\BibitemShut {NoStop}%
	\bibitem [{\citenamefont {Legrand}\ \emph {et~al.}(2017)\citenamefont
		{Legrand}, \citenamefont {Maccariello}, \citenamefont {Reyren}, \citenamefont
		{Garcia}, \citenamefont {Moutafis}, \citenamefont {Moreau-Luchaire},
		\citenamefont {Collin}, \citenamefont {Bouzehouane}, \citenamefont {Cros},\
		and\ \citenamefont {Fert}}]{Legrand2017}%
	\BibitemOpen
	\bibfield  {author} {\bibinfo {author} {\bibfnamefont {W.}~\bibnamefont
			{Legrand}}, \bibinfo {author} {\bibfnamefont {D.}~\bibnamefont
			{Maccariello}}, \bibinfo {author} {\bibfnamefont {N.}~\bibnamefont {Reyren}},
		\bibinfo {author} {\bibfnamefont {K.}~\bibnamefont {Garcia}}, \bibinfo
		{author} {\bibfnamefont {C.}~\bibnamefont {Moutafis}}, \bibinfo {author}
		{\bibfnamefont {C.}~\bibnamefont {Moreau-Luchaire}}, \bibinfo {author}
		{\bibfnamefont {S.}~\bibnamefont {Collin}}, \bibinfo {author} {\bibfnamefont
			{K.}~\bibnamefont {Bouzehouane}}, \bibinfo {author} {\bibfnamefont
			{V.}~\bibnamefont {Cros}}, \ and\ \bibinfo {author} {\bibfnamefont
			{A.}~\bibnamefont {Fert}},\ }\bibfield  {title} {\enquote {\bibinfo {title}
			{Room-temperature current-induced generation and motion of sub-100 nm
				skyrmions},}\ }\href {\doibase 10.1021/acs.nanolett.7b00649} {\bibfield
		{journal} {\bibinfo  {journal} {Nano Letters}\ }\textbf {\bibinfo {volume}
			{17}},\ \bibinfo {pages} {2703--2712} (\bibinfo {year} {2017})}\BibitemShut
	{NoStop}%
	\bibitem [{\citenamefont {B{\"u}ttner}\ \emph {et~al.}(2017)\citenamefont
		{B{\"u}ttner}, \citenamefont {Lemesh}, \citenamefont {Schneider},
		\citenamefont {Pfau}, \citenamefont {G{\"u}nther}, \citenamefont {Hessing},
		\citenamefont {Geilhufe}, \citenamefont {Caretta}, \citenamefont {Engel},
		\citenamefont {Kr{\"u}ger}, \citenamefont {Viefhaus}, \citenamefont
		{Eisebitt},\ and\ \citenamefont {Beach}}]{Butner2017}%
	\BibitemOpen
	\bibfield  {author} {\bibinfo {author} {\bibfnamefont {F.}~\bibnamefont
			{B{\"u}ttner}}, \bibinfo {author} {\bibfnamefont {I.}~\bibnamefont {Lemesh}},
		\bibinfo {author} {\bibfnamefont {M.}~\bibnamefont {Schneider}}, \bibinfo
		{author} {\bibfnamefont {B.}~\bibnamefont {Pfau}}, \bibinfo {author}
		{\bibfnamefont {C.~M.}\ \bibnamefont {G{\"u}nther}}, \bibinfo {author}
		{\bibfnamefont {P.}~\bibnamefont {Hessing}}, \bibinfo {author} {\bibfnamefont
			{J.}~\bibnamefont {Geilhufe}}, \bibinfo {author} {\bibfnamefont
			{L.}~\bibnamefont {Caretta}}, \bibinfo {author} {\bibfnamefont
			{D.}~\bibnamefont {Engel}}, \bibinfo {author} {\bibfnamefont
			{B.}~\bibnamefont {Kr{\"u}ger}}, \bibinfo {author} {\bibfnamefont
			{J.}~\bibnamefont {Viefhaus}}, \bibinfo {author} {\bibfnamefont
			{S.}~\bibnamefont {Eisebitt}}, \ and\ \bibinfo {author} {\bibfnamefont
			{G.~S.~D.}\ \bibnamefont {Beach}},\ }\bibfield  {title} {\enquote {\bibinfo
			{title} {Field-free deterministic ultrafast creation of magnetic skyrmions by
				spin--orbit torques},}\ }\href {\doibase 10.1038/nnano.2017.178} {\bibfield
		{journal} {\bibinfo  {journal} {Nature Nanotechnology}\ }\textbf {\bibinfo
			{volume} {12}},\ \bibinfo {pages} {1040--1044} (\bibinfo {year}
		{2017})}\BibitemShut {NoStop}%
	\bibitem [{\citenamefont {Soumyanarayanan}\ \emph {et~al.}(2017)\citenamefont
		{Soumyanarayanan}, \citenamefont {Raju}, \citenamefont {Gonzalez~Oyarce},
		\citenamefont {Tan}, \citenamefont {Im}, \citenamefont {Petrovi{\'c}},
		\citenamefont {Ho}, \citenamefont {Khoo}, \citenamefont {Tran}, \citenamefont
		{Gan}, \citenamefont {Ernult},\ and\ \citenamefont
		{Panagopoulos}}]{Soumyanarayanan2017}%
	\BibitemOpen
	\bibfield  {author} {\bibinfo {author} {\bibfnamefont {A.}~\bibnamefont
			{Soumyanarayanan}}, \bibinfo {author} {\bibfnamefont {M.}~\bibnamefont
			{Raju}}, \bibinfo {author} {\bibfnamefont {A.~L.}\ \bibnamefont
			{Gonzalez~Oyarce}}, \bibinfo {author} {\bibfnamefont {A.~K.~C.}\ \bibnamefont
			{Tan}}, \bibinfo {author} {\bibfnamefont {M.-Y.}\ \bibnamefont {Im}},
		\bibinfo {author} {\bibfnamefont {A.~P.}\ \bibnamefont {Petrovi{\'c}}},
		\bibinfo {author} {\bibfnamefont {P.}~\bibnamefont {Ho}}, \bibinfo {author}
		{\bibfnamefont {K.~H.}\ \bibnamefont {Khoo}}, \bibinfo {author}
		{\bibfnamefont {M.}~\bibnamefont {Tran}}, \bibinfo {author} {\bibfnamefont
			{C.~K.}\ \bibnamefont {Gan}}, \bibinfo {author} {\bibfnamefont
			{F.}~\bibnamefont {Ernult}}, \ and\ \bibinfo {author} {\bibfnamefont
			{C.}~\bibnamefont {Panagopoulos}},\ }\bibfield  {title} {\enquote {\bibinfo
			{title} {Tunable room-temperature magnetic skyrmions in
				\uppercase{I}r/\uppercase{F}e/\uppercase{C}o/\uppercase{P}t multilayers},}\
	}\href {\doibase 10.1038/nmat4934} {\bibfield  {journal} {\bibinfo  {journal}
			{Nature Materials}\ }\textbf {\bibinfo {volume} {16}},\ \bibinfo {pages}
		{898--904} (\bibinfo {year} {2017})}\BibitemShut {NoStop}%
	\bibitem [{\citenamefont {Finizio}\ \emph {et~al.}(2019)\citenamefont
		{Finizio}, \citenamefont {Zeissler}, \citenamefont {Wintz}, \citenamefont
		{Mayr}, \citenamefont {We{\ss}els}, \citenamefont {Huxtable}, \citenamefont
		{Burnell}, \citenamefont {Marrows},\ and\ \citenamefont
		{Raabe}}]{Finizio2019}%
	\BibitemOpen
	\bibfield  {author} {\bibinfo {author} {\bibfnamefont {S.}~\bibnamefont
			{Finizio}}, \bibinfo {author} {\bibfnamefont {K.}~\bibnamefont {Zeissler}},
		\bibinfo {author} {\bibfnamefont {S.}~\bibnamefont {Wintz}}, \bibinfo
		{author} {\bibfnamefont {S.}~\bibnamefont {Mayr}}, \bibinfo {author}
		{\bibfnamefont {T.}~\bibnamefont {We{\ss}els}}, \bibinfo {author}
		{\bibfnamefont {A.~J.}\ \bibnamefont {Huxtable}}, \bibinfo {author}
		{\bibfnamefont {G.}~\bibnamefont {Burnell}}, \bibinfo {author} {\bibfnamefont
			{C.~H.}\ \bibnamefont {Marrows}}, \ and\ \bibinfo {author} {\bibfnamefont
			{J.}~\bibnamefont {Raabe}},\ }\bibfield  {title} {\enquote {\bibinfo {title}
			{Deterministic field-free skyrmion nucleation at a nanoengineered injector
				device},}\ }\href {\doibase 10.1021/acs.nanolett.9b02840} {\bibfield
		{journal} {\bibinfo  {journal} {Nano Letters}\ }\textbf {\bibinfo {volume}
			{19}},\ \bibinfo {pages} {7246--7255} (\bibinfo {year} {2019})}\BibitemShut
	{NoStop}%
	\bibitem [{\citenamefont {Wang}\ \emph {et~al.}(2020)\citenamefont {Wang},
		\citenamefont {Guo}, \citenamefont {Zhou}, \citenamefont {Zhao},
		\citenamefont {Xu}, \citenamefont {Tomasello}, \citenamefont {Bai},
		\citenamefont {Dong}, \citenamefont {Je}, \citenamefont {Chao}, \citenamefont
		{Han}, \citenamefont {Lee}, \citenamefont {Lee}, \citenamefont {Yao},
		\citenamefont {Han}, \citenamefont {Song}, \citenamefont {Wu}, \citenamefont
		{Carpentieri}, \citenamefont {Finocchio}, \citenamefont {Im}, \citenamefont
		{Lin},\ and\ \citenamefont {Jiang}}]{Wang2020}%
	\BibitemOpen
	\bibfield  {author} {\bibinfo {author} {\bibfnamefont {Z.}~\bibnamefont
			{Wang}}, \bibinfo {author} {\bibfnamefont {M.}~\bibnamefont {Guo}}, \bibinfo
		{author} {\bibfnamefont {H.-A.}\ \bibnamefont {Zhou}}, \bibinfo {author}
		{\bibfnamefont {L.}~\bibnamefont {Zhao}}, \bibinfo {author} {\bibfnamefont
			{T.}~\bibnamefont {Xu}}, \bibinfo {author} {\bibfnamefont {R.}~\bibnamefont
			{Tomasello}}, \bibinfo {author} {\bibfnamefont {H.}~\bibnamefont {Bai}},
		\bibinfo {author} {\bibfnamefont {Y.}~\bibnamefont {Dong}}, \bibinfo {author}
		{\bibfnamefont {S.-G.}\ \bibnamefont {Je}}, \bibinfo {author} {\bibfnamefont
			{W.}~\bibnamefont {Chao}}, \bibinfo {author} {\bibfnamefont {H.-S.}\
			\bibnamefont {Han}}, \bibinfo {author} {\bibfnamefont {S.}~\bibnamefont
			{Lee}}, \bibinfo {author} {\bibfnamefont {K.-S.}\ \bibnamefont {Lee}},
		\bibinfo {author} {\bibfnamefont {Y.}~\bibnamefont {Yao}}, \bibinfo {author}
		{\bibfnamefont {W.}~\bibnamefont {Han}}, \bibinfo {author} {\bibfnamefont
			{C.}~\bibnamefont {Song}}, \bibinfo {author} {\bibfnamefont {H.}~\bibnamefont
			{Wu}}, \bibinfo {author} {\bibfnamefont {M.}~\bibnamefont {Carpentieri}},
		\bibinfo {author} {\bibfnamefont {G.}~\bibnamefont {Finocchio}}, \bibinfo
		{author} {\bibfnamefont {M.-Y.}\ \bibnamefont {Im}}, \bibinfo {author}
		{\bibfnamefont {S.-Z.}\ \bibnamefont {Lin}}, \ and\ \bibinfo {author}
		{\bibfnamefont {W.}~\bibnamefont {Jiang}},\ }\bibfield  {title} {\enquote
		{\bibinfo {title} {Thermal generation, manipulation and thermoelectric
				detection of skyrmions},}\ }\href {\doibase 10.1038/s41928-020-00489-2}
	{\bibfield  {journal} {\bibinfo  {journal} {Nature Electronics}\ }\textbf
		{\bibinfo {volume} {3}},\ \bibinfo {pages} {672--679} (\bibinfo {year}
		{2020})}\BibitemShut {NoStop}%
	\bibitem [{\citenamefont {Woo}\ \emph {et~al.}(2018)\citenamefont {Woo},
		\citenamefont {Song}, \citenamefont {Zhang}, \citenamefont {Ezawa},
		\citenamefont {Zhou}, \citenamefont {Liu}, \citenamefont {Weigand},
		\citenamefont {Finizio}, \citenamefont {Raabe}, \citenamefont {Park},
		\citenamefont {Lee}, \citenamefont {Choi}, \citenamefont {Min}, \citenamefont
		{Koo},\ and\ \citenamefont {Chang}}]{Woo2018}%
	\BibitemOpen
	\bibfield  {author} {\bibinfo {author} {\bibfnamefont {S.}~\bibnamefont
			{Woo}}, \bibinfo {author} {\bibfnamefont {K.~M.}\ \bibnamefont {Song}},
		\bibinfo {author} {\bibfnamefont {X.}~\bibnamefont {Zhang}}, \bibinfo
		{author} {\bibfnamefont {M.}~\bibnamefont {Ezawa}}, \bibinfo {author}
		{\bibfnamefont {Y.}~\bibnamefont {Zhou}}, \bibinfo {author} {\bibfnamefont
			{X.}~\bibnamefont {Liu}}, \bibinfo {author} {\bibfnamefont {M.}~\bibnamefont
			{Weigand}}, \bibinfo {author} {\bibfnamefont {S.}~\bibnamefont {Finizio}},
		\bibinfo {author} {\bibfnamefont {J.}~\bibnamefont {Raabe}}, \bibinfo
		{author} {\bibfnamefont {M.-C.}\ \bibnamefont {Park}}, \bibinfo {author}
		{\bibfnamefont {K.-Y.}\ \bibnamefont {Lee}}, \bibinfo {author} {\bibfnamefont
			{J.~W.}\ \bibnamefont {Choi}}, \bibinfo {author} {\bibfnamefont {B.-C.}\
			\bibnamefont {Min}}, \bibinfo {author} {\bibfnamefont {H.~C.}\ \bibnamefont
			{Koo}}, \ and\ \bibinfo {author} {\bibfnamefont {J.}~\bibnamefont {Chang}},\
	}\bibfield  {title} {\enquote {\bibinfo {title} {Deterministic creation and
				deletion of a single magnetic skyrmion observed by direct time-resolved x-ray
				microscopy},}\ }\href {\doibase 10.1038/s41928-018-0070-8} {\bibfield
		{journal} {\bibinfo  {journal} {Nature Electronics}\ }\textbf {\bibinfo
			{volume} {1}},\ \bibinfo {pages} {288--296} (\bibinfo {year}
		{2018})}\BibitemShut {NoStop}%
	\bibitem [{\citenamefont {Yang}\ \emph {et~al.}(2021)\citenamefont {Yang},
		\citenamefont {Moon}, \citenamefont {Ju}, \citenamefont {Kim}, \citenamefont
		{Kim}, \citenamefont {Kim}, \citenamefont {Tran}, \citenamefont {Hong},\ and\
		\citenamefont {Hwang}}]{Yang2021AdvancedMaterials}%
	\BibitemOpen
	\bibfield  {author} {\bibinfo {author} {\bibfnamefont {S.}~\bibnamefont
			{Yang}}, \bibinfo {author} {\bibfnamefont {K.-W.}\ \bibnamefont {Moon}},
		\bibinfo {author} {\bibfnamefont {T.-S.}\ \bibnamefont {Ju}}, \bibinfo
		{author} {\bibfnamefont {C.}~\bibnamefont {Kim}}, \bibinfo {author}
		{\bibfnamefont {H.-J.}\ \bibnamefont {Kim}}, \bibinfo {author} {\bibfnamefont
			{J.}~\bibnamefont {Kim}}, \bibinfo {author} {\bibfnamefont {B.~X.}\
			\bibnamefont {Tran}}, \bibinfo {author} {\bibfnamefont {J.-I.}\ \bibnamefont
			{Hong}}, \ and\ \bibinfo {author} {\bibfnamefont {C.}~\bibnamefont {Hwang}},\
	}\bibfield  {title} {\enquote {\bibinfo {title} {Electrical generation and
				deletion of magnetic skyrmion-bubbles via vertical current injection},}\
	}\href {\doibase https://doi.org/10.1002/adma.202104406} {\bibfield
		{journal} {\bibinfo  {journal} {Advanced Materials}\ }\textbf {\bibinfo
			{volume} {33}},\ \bibinfo {pages} {2104406} (\bibinfo {year}
		{2021})}\BibitemShut {NoStop}%
	\bibitem [{\citenamefont {Maccariello}\ \emph {et~al.}(2018)\citenamefont
		{Maccariello}, \citenamefont {Legrand}, \citenamefont {Reyren}, \citenamefont
		{Garcia}, \citenamefont {Bouzehouane}, \citenamefont {Collin}, \citenamefont
		{Cros},\ and\ \citenamefont {Fert}}]{Maccariello2018}%
	\BibitemOpen
	\bibfield  {author} {\bibinfo {author} {\bibfnamefont {D.}~\bibnamefont
			{Maccariello}}, \bibinfo {author} {\bibfnamefont {W.}~\bibnamefont
			{Legrand}}, \bibinfo {author} {\bibfnamefont {N.}~\bibnamefont {Reyren}},
		\bibinfo {author} {\bibfnamefont {K.}~\bibnamefont {Garcia}}, \bibinfo
		{author} {\bibfnamefont {K.}~\bibnamefont {Bouzehouane}}, \bibinfo {author}
		{\bibfnamefont {S.}~\bibnamefont {Collin}}, \bibinfo {author} {\bibfnamefont
			{V.}~\bibnamefont {Cros}}, \ and\ \bibinfo {author} {\bibfnamefont
			{A.}~\bibnamefont {Fert}},\ }\bibfield  {title} {\enquote {\bibinfo {title}
			{Electrical detection of single magnetic skyrmions in metallic multilayers at
				room temperature},}\ }\href {\doibase 10.1038/s41565-017-0044-4} {\bibfield
		{journal} {\bibinfo  {journal} {Nature Nanotechnology}\ }\textbf {\bibinfo
			{volume} {13}},\ \bibinfo {pages} {233--237} (\bibinfo {year}
		{2018})}\BibitemShut {NoStop}%
	\bibitem [{\citenamefont {Zeissler}\ \emph {et~al.}(2018)\citenamefont
		{Zeissler}, \citenamefont {Finizio}, \citenamefont {Shahbazi}, \citenamefont
		{Massey}, \citenamefont {Ma'Mari}, \citenamefont {Bracher}, \citenamefont
		{Kleibert}, \citenamefont {Rosamond}, \citenamefont {Linfield}, \citenamefont
		{Moore}, \citenamefont {Raabe}, \citenamefont {Burnell},\ and\ \citenamefont
		{Marrows}}]{Zeissler2018}%
	\BibitemOpen
	\bibfield  {author} {\bibinfo {author} {\bibfnamefont {K.}~\bibnamefont
			{Zeissler}}, \bibinfo {author} {\bibfnamefont {S.}~\bibnamefont {Finizio}},
		\bibinfo {author} {\bibfnamefont {K.}~\bibnamefont {Shahbazi}}, \bibinfo
		{author} {\bibfnamefont {J.}~\bibnamefont {Massey}}, \bibinfo {author}
		{\bibfnamefont {F.~A.}\ \bibnamefont {Ma'Mari}}, \bibinfo {author}
		{\bibfnamefont {D.~M.}\ \bibnamefont {Bracher}}, \bibinfo {author}
		{\bibfnamefont {A.}~\bibnamefont {Kleibert}}, \bibinfo {author}
		{\bibfnamefont {M.~C.}\ \bibnamefont {Rosamond}}, \bibinfo {author}
		{\bibfnamefont {E.~H.}\ \bibnamefont {Linfield}}, \bibinfo {author}
		{\bibfnamefont {T.~A.}\ \bibnamefont {Moore}}, \bibinfo {author}
		{\bibfnamefont {J.}~\bibnamefont {Raabe}}, \bibinfo {author} {\bibfnamefont
			{G.}~\bibnamefont {Burnell}}, \ and\ \bibinfo {author} {\bibfnamefont
			{C.~H.}\ \bibnamefont {Marrows}},\ }\bibfield  {title} {\enquote {\bibinfo
			{title} {Discrete hall resistivity contribution from n{\'e}el skyrmions in
				multilayer nanodiscs},}\ }\href {\doibase 10.1038/s41565-018-0268-y}
	{\bibfield  {journal} {\bibinfo  {journal} {Nature Nanotechnology}\ }\textbf
		{\bibinfo {volume} {13}},\ \bibinfo {pages} {1161--1166} (\bibinfo {year}
		{2018})}\BibitemShut {NoStop}%
	\bibitem [{\citenamefont {Hanneken}\ \emph {et~al.}(2015)\citenamefont
		{Hanneken}, \citenamefont {Otte}, \citenamefont {Kubetzka}, \citenamefont
		{Dup{\'e}}, \citenamefont {Romming}, \citenamefont {von Bergmann},
		\citenamefont {Wiesendanger},\ and\ \citenamefont {Heinze}}]{Hanneken2015}%
	\BibitemOpen
	\bibfield  {author} {\bibinfo {author} {\bibfnamefont {C.}~\bibnamefont
			{Hanneken}}, \bibinfo {author} {\bibfnamefont {F.}~\bibnamefont {Otte}},
		\bibinfo {author} {\bibfnamefont {A.}~\bibnamefont {Kubetzka}}, \bibinfo
		{author} {\bibfnamefont {B.}~\bibnamefont {Dup{\'e}}}, \bibinfo {author}
		{\bibfnamefont {N.}~\bibnamefont {Romming}}, \bibinfo {author} {\bibfnamefont
			{K.}~\bibnamefont {von Bergmann}}, \bibinfo {author} {\bibfnamefont
			{R.}~\bibnamefont {Wiesendanger}}, \ and\ \bibinfo {author} {\bibfnamefont
			{S.}~\bibnamefont {Heinze}},\ }\bibfield  {title} {\enquote {\bibinfo {title}
			{Electrical detection of magnetic skyrmions by tunnelling non-collinear
				magnetoresistance},}\ }\href {\doibase 10.1038/nnano.2015.218} {\bibfield
		{journal} {\bibinfo  {journal} {Nature Nanotechnology}\ }\textbf {\bibinfo
			{volume} {10}},\ \bibinfo {pages} {1039--1042} (\bibinfo {year}
		{2015})}\BibitemShut {NoStop}%
	\bibitem [{\citenamefont {Chen}\ \emph {et~al.}(2024)\citenamefont {Chen},
		\citenamefont {Lourembam}, \citenamefont {Ho}, \citenamefont {Toh},
		\citenamefont {Huang}, \citenamefont {Chen}, \citenamefont {Tan},
		\citenamefont {Yap}, \citenamefont {Lim}, \citenamefont {Tan}, \citenamefont
		{Suraj}, \citenamefont {Sim}, \citenamefont {Toh}, \citenamefont {Lim},
		\citenamefont {Lim}, \citenamefont {Zhou}, \citenamefont {Chung},
		\citenamefont {Lim},\ and\ \citenamefont {Soumyanarayanan}}]{Chen2024}%
	\BibitemOpen
	\bibfield  {author} {\bibinfo {author} {\bibfnamefont {S.}~\bibnamefont
			{Chen}}, \bibinfo {author} {\bibfnamefont {J.}~\bibnamefont {Lourembam}},
		\bibinfo {author} {\bibfnamefont {P.}~\bibnamefont {Ho}}, \bibinfo {author}
		{\bibfnamefont {A.~K.~J.}\ \bibnamefont {Toh}}, \bibinfo {author}
		{\bibfnamefont {J.}~\bibnamefont {Huang}}, \bibinfo {author} {\bibfnamefont
			{X.}~\bibnamefont {Chen}}, \bibinfo {author} {\bibfnamefont {H.~K.}\
			\bibnamefont {Tan}}, \bibinfo {author} {\bibfnamefont {S.~L.~K.}\
			\bibnamefont {Yap}}, \bibinfo {author} {\bibfnamefont {R.~J.~J.}\
			\bibnamefont {Lim}}, \bibinfo {author} {\bibfnamefont {H.~R.}\ \bibnamefont
			{Tan}}, \bibinfo {author} {\bibfnamefont {T.~S.}\ \bibnamefont {Suraj}},
		\bibinfo {author} {\bibfnamefont {M.~I.}\ \bibnamefont {Sim}}, \bibinfo
		{author} {\bibfnamefont {Y.~T.}\ \bibnamefont {Toh}}, \bibinfo {author}
		{\bibfnamefont {I.}~\bibnamefont {Lim}}, \bibinfo {author} {\bibfnamefont
			{N.~C.~B.}\ \bibnamefont {Lim}}, \bibinfo {author} {\bibfnamefont
			{J.}~\bibnamefont {Zhou}}, \bibinfo {author} {\bibfnamefont {H.~J.}\
			\bibnamefont {Chung}}, \bibinfo {author} {\bibfnamefont {S.~T.}\ \bibnamefont
			{Lim}}, \ and\ \bibinfo {author} {\bibfnamefont {A.}~\bibnamefont
			{Soumyanarayanan}},\ }\bibfield  {title} {\enquote {\bibinfo {title}
			{All-electrical skyrmionic magnetic tunnel junction},}\ }\href {\doibase
		10.1038/s41586-024-07131-7} {\bibfield  {journal} {\bibinfo  {journal}
			{Nature}\ }\textbf {\bibinfo {volume} {627}},\ \bibinfo {pages} {522--527}
		(\bibinfo {year} {2024})}\BibitemShut {NoStop}%
	\bibitem [{\citenamefont {Urrestarazu~Larra{\~n}aga}\ \emph
		{et~al.}(2024)\citenamefont {Urrestarazu~Larra{\~n}aga}, \citenamefont
		{Sisodia}, \citenamefont {Guedas}, \citenamefont {Pham}, \citenamefont
		{Di~Manici}, \citenamefont {Masseboeuf}, \citenamefont {Garello},
		\citenamefont {Disdier}, \citenamefont {Fernandez}, \citenamefont {Wintz},
		\citenamefont {Weigand}, \citenamefont {Belmeguenai}, \citenamefont
		{Pizzini}, \citenamefont {Sousa}, \citenamefont {Buda-Prejbeanu},
		\citenamefont {Gaudin},\ and\ \citenamefont {Boulle}}]{Urrestarazu2024}%
	\BibitemOpen
	\bibfield  {author} {\bibinfo {author} {\bibfnamefont {J.}~\bibnamefont
			{Urrestarazu~Larra{\~n}aga}}, \bibinfo {author} {\bibfnamefont
			{N.}~\bibnamefont {Sisodia}}, \bibinfo {author} {\bibfnamefont
			{R.}~\bibnamefont {Guedas}}, \bibinfo {author} {\bibfnamefont {V.~T.}\
			\bibnamefont {Pham}}, \bibinfo {author} {\bibfnamefont {I.}~\bibnamefont
			{Di~Manici}}, \bibinfo {author} {\bibfnamefont {A.}~\bibnamefont
			{Masseboeuf}}, \bibinfo {author} {\bibfnamefont {K.}~\bibnamefont {Garello}},
		\bibinfo {author} {\bibfnamefont {F.}~\bibnamefont {Disdier}}, \bibinfo
		{author} {\bibfnamefont {B.}~\bibnamefont {Fernandez}}, \bibinfo {author}
		{\bibfnamefont {S.}~\bibnamefont {Wintz}}, \bibinfo {author} {\bibfnamefont
			{M.}~\bibnamefont {Weigand}}, \bibinfo {author} {\bibfnamefont
			{M.}~\bibnamefont {Belmeguenai}}, \bibinfo {author} {\bibfnamefont
			{S.}~\bibnamefont {Pizzini}}, \bibinfo {author} {\bibfnamefont {R.~C.}\
			\bibnamefont {Sousa}}, \bibinfo {author} {\bibfnamefont {L.~D.}\ \bibnamefont
			{Buda-Prejbeanu}}, \bibinfo {author} {\bibfnamefont {G.}~\bibnamefont
			{Gaudin}}, \ and\ \bibinfo {author} {\bibfnamefont {O.}~\bibnamefont
			{Boulle}},\ }\bibfield  {title} {\enquote {\bibinfo {title} {Electrical
				detection and nucleation of a magnetic skyrmion in a magnetic tunnel junction
				observed via operando magnetic microscopy},}\ }\href {\doibase
		10.1021/acs.nanolett.4c00316} {\bibfield  {journal} {\bibinfo  {journal}
			{Nano Letters}\ }\textbf {\bibinfo {volume} {24}},\ \bibinfo {pages}
		{3557--3565} (\bibinfo {year} {2024})}\BibitemShut {NoStop}%
	\bibitem [{\citenamefont {Bourianoff}\ \emph {et~al.}(2018)\citenamefont
		{Bourianoff}, \citenamefont {Pinna}, \citenamefont {Sitte},\ and\
		\citenamefont {Everschor-Sitte}}]{Bourianoff2018}%
	\BibitemOpen
	\bibfield  {author} {\bibinfo {author} {\bibfnamefont {G.}~\bibnamefont
			{Bourianoff}}, \bibinfo {author} {\bibfnamefont {D.}~\bibnamefont {Pinna}},
		\bibinfo {author} {\bibfnamefont {M.}~\bibnamefont {Sitte}}, \ and\ \bibinfo
		{author} {\bibfnamefont {K.}~\bibnamefont {Everschor-Sitte}},\ }\bibfield
	{title} {\enquote {\bibinfo {title} {Potential implementation of reservoir
				computing models based on magnetic skyrmions},}\ }\href {\doibase
		10.1063/1.5006918} {\bibfield  {journal} {\bibinfo  {journal} {AIP Advances}\
		}\textbf {\bibinfo {volume} {8}},\ \bibinfo {pages} {055602} (\bibinfo {year}
		{2018})}\BibitemShut {NoStop}%
	\bibitem [{\citenamefont {Song}\ \emph {et~al.}(2020)\citenamefont {Song},
		\citenamefont {Jeong}, \citenamefont {Pan}, \citenamefont {Zhang},
		\citenamefont {Xia}, \citenamefont {Cha}, \citenamefont {Park}, \citenamefont
		{Kim}, \citenamefont {Finizio}, \citenamefont {Raabe}, \citenamefont {Chang},
		\citenamefont {Zhou}, \citenamefont {Zhao}, \citenamefont {Kang},
		\citenamefont {Ju},\ and\ \citenamefont {Woo}}]{Song2020}%
	\BibitemOpen
	\bibfield  {author} {\bibinfo {author} {\bibfnamefont {K.~M.}\ \bibnamefont
			{Song}}, \bibinfo {author} {\bibfnamefont {J.-S.}\ \bibnamefont {Jeong}},
		\bibinfo {author} {\bibfnamefont {B.}~\bibnamefont {Pan}}, \bibinfo {author}
		{\bibfnamefont {X.}~\bibnamefont {Zhang}}, \bibinfo {author} {\bibfnamefont
			{J.}~\bibnamefont {Xia}}, \bibinfo {author} {\bibfnamefont {S.}~\bibnamefont
			{Cha}}, \bibinfo {author} {\bibfnamefont {T.-E.}\ \bibnamefont {Park}},
		\bibinfo {author} {\bibfnamefont {K.}~\bibnamefont {Kim}}, \bibinfo {author}
		{\bibfnamefont {S.}~\bibnamefont {Finizio}}, \bibinfo {author} {\bibfnamefont
			{J.}~\bibnamefont {Raabe}}, \bibinfo {author} {\bibfnamefont
			{J.}~\bibnamefont {Chang}}, \bibinfo {author} {\bibfnamefont
			{Y.}~\bibnamefont {Zhou}}, \bibinfo {author} {\bibfnamefont {W.}~\bibnamefont
			{Zhao}}, \bibinfo {author} {\bibfnamefont {W.}~\bibnamefont {Kang}}, \bibinfo
		{author} {\bibfnamefont {H.}~\bibnamefont {Ju}}, \ and\ \bibinfo {author}
		{\bibfnamefont {S.}~\bibnamefont {Woo}},\ }\bibfield  {title} {\enquote
		{\bibinfo {title} {Skyrmion-based artificial synapses for neuromorphic
				computing},}\ }\href {\doibase 10.1038/s41928-020-0385-0} {\bibfield
		{journal} {\bibinfo  {journal} {Nature Electronics}\ }\textbf {\bibinfo
			{volume} {3}},\ \bibinfo {pages} {148--155} (\bibinfo {year}
		{2020})}\BibitemShut {NoStop}%
	\bibitem [{\citenamefont {Grollier}\ \emph {et~al.}(2020)\citenamefont
		{Grollier}, \citenamefont {Querlioz}, \citenamefont {Camsari}, \citenamefont
		{Everschor-Sitte}, \citenamefont {Fukami},\ and\ \citenamefont
		{Stiles}}]{Grollier2020}%
	\BibitemOpen
	\bibfield  {author} {\bibinfo {author} {\bibfnamefont {J.}~\bibnamefont
			{Grollier}}, \bibinfo {author} {\bibfnamefont {D.}~\bibnamefont {Querlioz}},
		\bibinfo {author} {\bibfnamefont {K.~Y.}\ \bibnamefont {Camsari}}, \bibinfo
		{author} {\bibfnamefont {K.}~\bibnamefont {Everschor-Sitte}}, \bibinfo
		{author} {\bibfnamefont {S.}~\bibnamefont {Fukami}}, \ and\ \bibinfo {author}
		{\bibfnamefont {M.~D.}\ \bibnamefont {Stiles}},\ }\bibfield  {title}
	{\enquote {\bibinfo {title} {Neuromorphic spintronics},}\ }\href {\doibase
		10.1038/s41928-019-0360-9} {\bibfield  {journal} {\bibinfo  {journal} {Nature
				Electronics}\ }\textbf {\bibinfo {volume} {3}},\ \bibinfo {pages} {360--370}
		(\bibinfo {year} {2020})}\BibitemShut {NoStop}%
	\bibitem [{\citenamefont {Huang}\ \emph {et~al.}(2017)\citenamefont {Huang},
		\citenamefont {Kang}, \citenamefont {Zhang}, \citenamefont {Zhou},\ and\
		\citenamefont {Zhao}}]{Huang2017}%
	\BibitemOpen
	\bibfield  {author} {\bibinfo {author} {\bibfnamefont {Y.}~\bibnamefont
			{Huang}}, \bibinfo {author} {\bibfnamefont {W.}~\bibnamefont {Kang}},
		\bibinfo {author} {\bibfnamefont {X.}~\bibnamefont {Zhang}}, \bibinfo
		{author} {\bibfnamefont {Y.}~\bibnamefont {Zhou}}, \ and\ \bibinfo {author}
		{\bibfnamefont {W.}~\bibnamefont {Zhao}},\ }\bibfield  {title} {\enquote
		{\bibinfo {title} {Magnetic skyrmion-based synaptic devices},}\ }\href
	{\doibase 10.1088/1361-6528/aa5838} {\bibfield  {journal} {\bibinfo
			{journal} {Nanotechnology}\ }\textbf {\bibinfo {volume} {28}},\ \bibinfo
		{pages} {08LT02} (\bibinfo {year} {2017})}\BibitemShut {NoStop}%
	\bibitem [{\citenamefont {Sharad}\ \emph {et~al.}(2012)\citenamefont {Sharad},
		\citenamefont {Augustine}, \citenamefont {Panagopoulos},\ and\ \citenamefont
		{Roy}}]{Sharad2012}%
	\BibitemOpen
	\bibfield  {author} {\bibinfo {author} {\bibfnamefont {M.}~\bibnamefont
			{Sharad}}, \bibinfo {author} {\bibfnamefont {C.}~\bibnamefont {Augustine}},
		\bibinfo {author} {\bibfnamefont {G.}~\bibnamefont {Panagopoulos}}, \ and\
		\bibinfo {author} {\bibfnamefont {K.}~\bibnamefont {Roy}},\ }\bibfield
	{title} {\enquote {\bibinfo {title} {Spin-based neuron model with domain-wall
				magnets as synapse},}\ }\href {\doibase 10.1109/TNANO.2012.2202125}
	{\bibfield  {journal} {\bibinfo  {journal} {IEEE Transactions on
				Nanotechnology}\ }\textbf {\bibinfo {volume} {11}},\ \bibinfo {pages}
		{843--853} (\bibinfo {year} {2012})}\BibitemShut {NoStop}%
	\bibitem [{\citenamefont {Chen}\ \emph {et~al.}(2020)\citenamefont {Chen},
		\citenamefont {Li}, \citenamefont {Li}, \citenamefont {Miles}, \citenamefont
		{Indiveri}, \citenamefont {Furber}, \citenamefont {Pavlidis},\ and\
		\citenamefont {Moutafis}}]{Chen2020}%
	\BibitemOpen
	\bibfield  {author} {\bibinfo {author} {\bibfnamefont {R.}~\bibnamefont
			{Chen}}, \bibinfo {author} {\bibfnamefont {C.}~\bibnamefont {Li}}, \bibinfo
		{author} {\bibfnamefont {Y.}~\bibnamefont {Li}}, \bibinfo {author}
		{\bibfnamefont {J.~J.}\ \bibnamefont {Miles}}, \bibinfo {author}
		{\bibfnamefont {G.}~\bibnamefont {Indiveri}}, \bibinfo {author}
		{\bibfnamefont {S.}~\bibnamefont {Furber}}, \bibinfo {author} {\bibfnamefont
			{V.~F.}\ \bibnamefont {Pavlidis}}, \ and\ \bibinfo {author} {\bibfnamefont
			{C.}~\bibnamefont {Moutafis}},\ }\bibfield  {title} {\enquote {\bibinfo
			{title} {Nanoscale room-temperature multilayer skyrmionic synapse for deep
				spiking neural networks},}\ }\href {\doibase
		10.1103/PhysRevApplied.14.014096} {\bibfield  {journal} {\bibinfo  {journal}
			{Phys. Rev. Appl.}\ }\textbf {\bibinfo {volume} {14}},\ \bibinfo {pages}
		{014096} (\bibinfo {year} {2020})}\BibitemShut {NoStop}%
	\bibitem [{\citenamefont {Pinna}\ \emph {et~al.}(2018)\citenamefont {Pinna},
		\citenamefont {Abreu~Araujo}, \citenamefont {Kim}, \citenamefont {Cros},
		\citenamefont {Querlioz}, \citenamefont {Bessiere}, \citenamefont {Droulez},\
		and\ \citenamefont {Grollier}}]{Pinna2018}%
	\BibitemOpen
	\bibfield  {author} {\bibinfo {author} {\bibfnamefont {D.}~\bibnamefont
			{Pinna}}, \bibinfo {author} {\bibfnamefont {F.}~\bibnamefont {Abreu~Araujo}},
		\bibinfo {author} {\bibfnamefont {J.-V.}\ \bibnamefont {Kim}}, \bibinfo
		{author} {\bibfnamefont {V.}~\bibnamefont {Cros}}, \bibinfo {author}
		{\bibfnamefont {D.}~\bibnamefont {Querlioz}}, \bibinfo {author}
		{\bibfnamefont {P.}~\bibnamefont {Bessiere}}, \bibinfo {author}
		{\bibfnamefont {J.}~\bibnamefont {Droulez}}, \ and\ \bibinfo {author}
		{\bibfnamefont {J.}~\bibnamefont {Grollier}},\ }\bibfield  {title} {\enquote
		{\bibinfo {title} {Skyrmion gas manipulation for probabilistic computing},}\
	}\href {\doibase 10.1103/PhysRevApplied.9.064018} {\bibfield  {journal}
		{\bibinfo  {journal} {Phys. Rev. Appl.}\ }\textbf {\bibinfo {volume} {9}},\
		\bibinfo {pages} {064018} (\bibinfo {year} {2018})}\BibitemShut {NoStop}%
	\bibitem [{\citenamefont {Z{\'a}zvorka}\ \emph {et~al.}(2019)\citenamefont
		{Z{\'a}zvorka}, \citenamefont {Jakobs}, \citenamefont {Heinze}, \citenamefont
		{Keil}, \citenamefont {Kromin}, \citenamefont {Jaiswal}, \citenamefont
		{Litzius}, \citenamefont {Jakob}, \citenamefont {Virnau}, \citenamefont
		{Pinna}, \citenamefont {Everschor-Sitte}, \citenamefont {R{\'o}zsa},
		\citenamefont {Donges}, \citenamefont {Nowak},\ and\ \citenamefont
		{Kl{\"a}ui}}]{Zazvorka2019}%
	\BibitemOpen
	\bibfield  {author} {\bibinfo {author} {\bibfnamefont {J.}~\bibnamefont
			{Z{\'a}zvorka}}, \bibinfo {author} {\bibfnamefont {F.}~\bibnamefont
			{Jakobs}}, \bibinfo {author} {\bibfnamefont {D.}~\bibnamefont {Heinze}},
		\bibinfo {author} {\bibfnamefont {N.}~\bibnamefont {Keil}}, \bibinfo {author}
		{\bibfnamefont {S.}~\bibnamefont {Kromin}}, \bibinfo {author} {\bibfnamefont
			{S.}~\bibnamefont {Jaiswal}}, \bibinfo {author} {\bibfnamefont
			{K.}~\bibnamefont {Litzius}}, \bibinfo {author} {\bibfnamefont
			{G.}~\bibnamefont {Jakob}}, \bibinfo {author} {\bibfnamefont
			{P.}~\bibnamefont {Virnau}}, \bibinfo {author} {\bibfnamefont
			{D.}~\bibnamefont {Pinna}}, \bibinfo {author} {\bibfnamefont
			{K.}~\bibnamefont {Everschor-Sitte}}, \bibinfo {author} {\bibfnamefont
			{L.}~\bibnamefont {R{\'o}zsa}}, \bibinfo {author} {\bibfnamefont
			{A.}~\bibnamefont {Donges}}, \bibinfo {author} {\bibfnamefont
			{U.}~\bibnamefont {Nowak}}, \ and\ \bibinfo {author} {\bibfnamefont
			{M.}~\bibnamefont {Kl{\"a}ui}},\ }\bibfield  {title} {\enquote {\bibinfo
			{title} {Thermal skyrmion diffusion used in a reshuffler device},}\ }\href
	{\doibase 10.1038/s41565-019-0436-8} {\bibfield  {journal} {\bibinfo
			{journal} {Nature Nanotechnology}\ }\textbf {\bibinfo {volume} {14}},\
		\bibinfo {pages} {658--661} (\bibinfo {year} {2019})}\BibitemShut {NoStop}%
	\bibitem [{\citenamefont {Prychynenko}\ \emph {et~al.}(2018)\citenamefont
		{Prychynenko}, \citenamefont {Sitte}, \citenamefont {Litzius}, \citenamefont
		{Kr\"uger}, \citenamefont {Bourianoff}, \citenamefont {Kl\"aui},
		\citenamefont {Sinova},\ and\ \citenamefont
		{Everschor-Sitte}}]{Prychynenko2018}%
	\BibitemOpen
	\bibfield  {author} {\bibinfo {author} {\bibfnamefont {D.}~\bibnamefont
			{Prychynenko}}, \bibinfo {author} {\bibfnamefont {M.}~\bibnamefont {Sitte}},
		\bibinfo {author} {\bibfnamefont {K.}~\bibnamefont {Litzius}}, \bibinfo
		{author} {\bibfnamefont {B.}~\bibnamefont {Kr\"uger}}, \bibinfo {author}
		{\bibfnamefont {G.}~\bibnamefont {Bourianoff}}, \bibinfo {author}
		{\bibfnamefont {M.}~\bibnamefont {Kl\"aui}}, \bibinfo {author} {\bibfnamefont
			{J.}~\bibnamefont {Sinova}}, \ and\ \bibinfo {author} {\bibfnamefont
			{K.}~\bibnamefont {Everschor-Sitte}},\ }\bibfield  {title} {\enquote
		{\bibinfo {title} {Magnetic skyrmion as a nonlinear resistive element: A
				potential building block for reservoir computing},}\ }\href {\doibase
		10.1103/PhysRevApplied.9.014034} {\bibfield  {journal} {\bibinfo  {journal}
			{Phys. Rev. Appl.}\ }\textbf {\bibinfo {volume} {9}},\ \bibinfo {pages}
		{014034} (\bibinfo {year} {2018})}\BibitemShut {NoStop}%
	\bibitem [{\citenamefont {Raab}\ \emph {et~al.}(2022)\citenamefont {Raab},
		\citenamefont {Brems}, \citenamefont {Beneke}, \citenamefont {Dohi},
		\citenamefont {Roth{\"o}rl}, \citenamefont {Kammerbauer}, \citenamefont
		{Mentink},\ and\ \citenamefont {Kl{\"a}ui}}]{Raab2022}%
	\BibitemOpen
	\bibfield  {author} {\bibinfo {author} {\bibfnamefont {K.}~\bibnamefont
			{Raab}}, \bibinfo {author} {\bibfnamefont {M.~A.}\ \bibnamefont {Brems}},
		\bibinfo {author} {\bibfnamefont {G.}~\bibnamefont {Beneke}}, \bibinfo
		{author} {\bibfnamefont {T.}~\bibnamefont {Dohi}}, \bibinfo {author}
		{\bibfnamefont {J.}~\bibnamefont {Roth{\"o}rl}}, \bibinfo {author}
		{\bibfnamefont {F.}~\bibnamefont {Kammerbauer}}, \bibinfo {author}
		{\bibfnamefont {J.~H.}\ \bibnamefont {Mentink}}, \ and\ \bibinfo {author}
		{\bibfnamefont {M.}~\bibnamefont {Kl{\"a}ui}},\ }\bibfield  {title} {\enquote
		{\bibinfo {title} {Brownian reservoir computing realized using geometrically
				confined skyrmion dynamics},}\ }\href {\doibase 10.1038/s41467-022-34309-2}
	{\bibfield  {journal} {\bibinfo  {journal} {Nature Communications}\ }\textbf
		{\bibinfo {volume} {13}},\ \bibinfo {pages} {6982} (\bibinfo {year}
		{2022})}\BibitemShut {NoStop}%
	\bibitem [{\citenamefont {Yokouchi}\ \emph {et~al.}(2023)\citenamefont
		{Yokouchi}, \citenamefont {Sugimoto}, \citenamefont {Rana}, \citenamefont
		{Seki}, \citenamefont {Ogawa}, \citenamefont {Shiomi}, \citenamefont
		{Kasai},\ and\ \citenamefont {Otani}}]{Yokouchi2023}%
	\BibitemOpen
	\bibfield  {author} {\bibinfo {author} {\bibfnamefont {T.}~\bibnamefont
			{Yokouchi}}, \bibinfo {author} {\bibfnamefont {S.}~\bibnamefont {Sugimoto}},
		\bibinfo {author} {\bibfnamefont {B.}~\bibnamefont {Rana}}, \bibinfo {author}
		{\bibfnamefont {S.}~\bibnamefont {Seki}}, \bibinfo {author} {\bibfnamefont
			{N.}~\bibnamefont {Ogawa}}, \bibinfo {author} {\bibfnamefont
			{Y.}~\bibnamefont {Shiomi}}, \bibinfo {author} {\bibfnamefont
			{S.}~\bibnamefont {Kasai}}, \ and\ \bibinfo {author} {\bibfnamefont
			{Y.}~\bibnamefont {Otani}},\ }\bibfield  {title} {\enquote {\bibinfo {title}
			{Pattern recognition with neuromorphic computing using magnetic
				field--induced dynamics of skyrmions},}\ }\href {\doibase
		10.1126/sciadv.abq5652} {\bibfield  {journal} {\bibinfo  {journal} {Science
				Advances}\ }\textbf {\bibinfo {volume} {8}},\ \bibinfo {pages} {eabq5652}
		(\bibinfo {year} {2023})}\BibitemShut {NoStop}%
	\bibitem [{\citenamefont {Sun}\ \emph {et~al.}(2023)\citenamefont {Sun},
		\citenamefont {Lin}, \citenamefont {Lei}, \citenamefont {Chen}, \citenamefont
		{Kang}, \citenamefont {Zhao}, \citenamefont {Wei}, \citenamefont {Chen},
		\citenamefont {Pang}, \citenamefont {Hu}, \citenamefont {Yang}, \citenamefont
		{Dong}, \citenamefont {Zhao}, \citenamefont {Liu}, \citenamefont {Yuan},
		\citenamefont {Ullrich}, \citenamefont {Back}, \citenamefont {Zhang},
		\citenamefont {Pan}, \citenamefont {Zhao}, \citenamefont {Feng},
		\citenamefont {Fert},\ and\ \citenamefont {Zhao}}]{Sun2023}%
	\BibitemOpen
	\bibfield  {author} {\bibinfo {author} {\bibfnamefont {Y.}~\bibnamefont
			{Sun}}, \bibinfo {author} {\bibfnamefont {T.}~\bibnamefont {Lin}}, \bibinfo
		{author} {\bibfnamefont {N.}~\bibnamefont {Lei}}, \bibinfo {author}
		{\bibfnamefont {X.}~\bibnamefont {Chen}}, \bibinfo {author} {\bibfnamefont
			{W.}~\bibnamefont {Kang}}, \bibinfo {author} {\bibfnamefont {Z.}~\bibnamefont
			{Zhao}}, \bibinfo {author} {\bibfnamefont {D.}~\bibnamefont {Wei}}, \bibinfo
		{author} {\bibfnamefont {C.}~\bibnamefont {Chen}}, \bibinfo {author}
		{\bibfnamefont {S.}~\bibnamefont {Pang}}, \bibinfo {author} {\bibfnamefont
			{L.}~\bibnamefont {Hu}}, \bibinfo {author} {\bibfnamefont {L.}~\bibnamefont
			{Yang}}, \bibinfo {author} {\bibfnamefont {E.}~\bibnamefont {Dong}}, \bibinfo
		{author} {\bibfnamefont {L.}~\bibnamefont {Zhao}}, \bibinfo {author}
		{\bibfnamefont {L.}~\bibnamefont {Liu}}, \bibinfo {author} {\bibfnamefont
			{Z.}~\bibnamefont {Yuan}}, \bibinfo {author} {\bibfnamefont {A.}~\bibnamefont
			{Ullrich}}, \bibinfo {author} {\bibfnamefont {C.~H.}\ \bibnamefont {Back}},
		\bibinfo {author} {\bibfnamefont {J.}~\bibnamefont {Zhang}}, \bibinfo
		{author} {\bibfnamefont {D.}~\bibnamefont {Pan}}, \bibinfo {author}
		{\bibfnamefont {J.}~\bibnamefont {Zhao}}, \bibinfo {author} {\bibfnamefont
			{M.}~\bibnamefont {Feng}}, \bibinfo {author} {\bibfnamefont {A.}~\bibnamefont
			{Fert}}, \ and\ \bibinfo {author} {\bibfnamefont {W.}~\bibnamefont {Zhao}},\
	}\bibfield  {title} {\enquote {\bibinfo {title} {Experimental demonstration
				of a skyrmion-enhanced strain-mediated physical reservoir computing
				system},}\ }\href@noop {} {\bibfield  {journal} {\bibinfo  {journal} {Nature
				Communications}\ }\textbf {\bibinfo {volume} {14}},\ \bibinfo {pages} {3434}
		(\bibinfo {year} {2023})}\BibitemShut {NoStop}%
	\bibitem [{\citenamefont {Frenkel}\ \emph {et~al.}(2019)\citenamefont
		{Frenkel}, \citenamefont {Lefebvre}, \citenamefont {Legat},\ and\
		\citenamefont {Bol}}]{Frenkel2019}%
	\BibitemOpen
	\bibfield  {author} {\bibinfo {author} {\bibfnamefont {C.}~\bibnamefont
			{Frenkel}}, \bibinfo {author} {\bibfnamefont {M.}~\bibnamefont {Lefebvre}},
		\bibinfo {author} {\bibfnamefont {J.-D.}\ \bibnamefont {Legat}}, \ and\
		\bibinfo {author} {\bibfnamefont {D.}~\bibnamefont {Bol}},\ }\bibfield
	{title} {\enquote {\bibinfo {title} {A 0.086-mm$^2$ 12.7-pj/sop 64k-synapse
				256-neuron online-learning digital spiking neuromorphic processor in 28-nm
				cmos},}\ }\href {\doibase 10.1109/TBCAS.2018.2880425} {\bibfield  {journal}
		{\bibinfo  {journal} {IEEE Transactions on Biomedical Circuits and Systems}\
		}\textbf {\bibinfo {volume} {13}},\ \bibinfo {pages} {145--158} (\bibinfo
		{year} {2019})}\BibitemShut {NoStop}%
	\bibitem [{\citenamefont {Attwell}\ and\ \citenamefont
		{Laughlin}(2001)}]{Attwell2001}%
	\BibitemOpen
	\bibfield  {author} {\bibinfo {author} {\bibfnamefont {D.}~\bibnamefont
			{Attwell}}\ and\ \bibinfo {author} {\bibfnamefont {S.~B.}\ \bibnamefont
			{Laughlin}},\ }\bibfield  {title} {\enquote {\bibinfo {title} {An energy
				budget for signaling in the grey matter of the brain},}\ }\href {\doibase
		10.1097/00004647-200110000-00001} {\bibfield  {journal} {\bibinfo  {journal}
			{Journal of Cerebral Blood Flow \& Metabolism}\ }\textbf {\bibinfo {volume}
			{21}},\ \bibinfo {pages} {1133--1145} (\bibinfo {year} {2001})}\BibitemShut
	{NoStop}%
	\bibitem [{\citenamefont {Harris}, \citenamefont {Jolivet},\ and\ \citenamefont
		{Attwell}(2012)}]{Harris2012}%
	\BibitemOpen
	\bibfield  {author} {\bibinfo {author} {\bibfnamefont {J.~J.}\ \bibnamefont
			{Harris}}, \bibinfo {author} {\bibfnamefont {R.}~\bibnamefont {Jolivet}}, \
		and\ \bibinfo {author} {\bibfnamefont {D.}~\bibnamefont {Attwell}},\
	}\bibfield  {title} {\enquote {\bibinfo {title} {Synaptic energy use and
				supply},}\ }\href {\doibase 10.1016/j.neuron.2012.08.019} {\bibfield
		{journal} {\bibinfo  {journal} {Neuron}\ }\textbf {\bibinfo {volume} {75}},\
		\bibinfo {pages} {762--777} (\bibinfo {year} {2012})}\BibitemShut {NoStop}%
	\bibitem [{\citenamefont {Chanaday}\ \emph {et~al.}(2019)\citenamefont
		{Chanaday}, \citenamefont {Cousin}, \citenamefont {Milosevic}, \citenamefont
		{Watanabe},\ and\ \citenamefont {Morgan}}]{Chanaday2019}%
	\BibitemOpen
	\bibfield  {author} {\bibinfo {author} {\bibfnamefont {N.~L.}\ \bibnamefont
			{Chanaday}}, \bibinfo {author} {\bibfnamefont {M.~A.}\ \bibnamefont
			{Cousin}}, \bibinfo {author} {\bibfnamefont {I.}~\bibnamefont {Milosevic}},
		\bibinfo {author} {\bibfnamefont {S.}~\bibnamefont {Watanabe}}, \ and\
		\bibinfo {author} {\bibfnamefont {J.~R.}\ \bibnamefont {Morgan}},\ }\bibfield
	{title} {\enquote {\bibinfo {title} {The synaptic vesicle cycle revisited:
				New insights into the modes and mechanisms},}\ }\href {\doibase
		10.1523/JNEUROSCI.1158-19.2019} {\bibfield  {journal} {\bibinfo  {journal}
			{Journal of Neuroscience}\ }\textbf {\bibinfo {volume} {39}},\ \bibinfo
		{pages} {8209--8216} (\bibinfo {year} {2019})}\BibitemShut {NoStop}%
	\bibitem [{\citenamefont {Kent}\ \emph {et~al.}(2021)\citenamefont {Kent},
		\citenamefont {Reynolds}, \citenamefont {Raftrey}, \citenamefont {Campbell},
		\citenamefont {Virasawmy}, \citenamefont {Dhuey}, \citenamefont {Chopdekar},
		\citenamefont {Hierro-Rodriguez}, \citenamefont {Sorrentino}, \citenamefont
		{Pereiro}, \citenamefont {Ferrer}, \citenamefont {Hellman}, \citenamefont
		{Sutcliffe},\ and\ \citenamefont {Fischer}}]{Kent2021}%
	\BibitemOpen
	\bibfield  {author} {\bibinfo {author} {\bibfnamefont {N.}~\bibnamefont
			{Kent}}, \bibinfo {author} {\bibfnamefont {N.}~\bibnamefont {Reynolds}},
		\bibinfo {author} {\bibfnamefont {D.}~\bibnamefont {Raftrey}}, \bibinfo
		{author} {\bibfnamefont {I.~T.~G.}\ \bibnamefont {Campbell}}, \bibinfo
		{author} {\bibfnamefont {S.}~\bibnamefont {Virasawmy}}, \bibinfo {author}
		{\bibfnamefont {S.}~\bibnamefont {Dhuey}}, \bibinfo {author} {\bibfnamefont
			{R.~V.}\ \bibnamefont {Chopdekar}}, \bibinfo {author} {\bibfnamefont
			{A.}~\bibnamefont {Hierro-Rodriguez}}, \bibinfo {author} {\bibfnamefont
			{A.}~\bibnamefont {Sorrentino}}, \bibinfo {author} {\bibfnamefont
			{E.}~\bibnamefont {Pereiro}}, \bibinfo {author} {\bibfnamefont
			{S.}~\bibnamefont {Ferrer}}, \bibinfo {author} {\bibfnamefont
			{F.}~\bibnamefont {Hellman}}, \bibinfo {author} {\bibfnamefont
			{P.}~\bibnamefont {Sutcliffe}}, \ and\ \bibinfo {author} {\bibfnamefont
			{P.}~\bibnamefont {Fischer}},\ }\bibfield  {title} {\enquote {\bibinfo
			{title} {Creation and observation of hopfions in magnetic multilayer
				systems},}\ }\href {\doibase 10.1038/s41467-021-21846-5} {\bibfield
		{journal} {\bibinfo  {journal} {Nature Communications}\ }\textbf {\bibinfo
			{volume} {12}},\ \bibinfo {pages} {1562} (\bibinfo {year}
		{2021})}\BibitemShut {NoStop}%
	\bibitem [{\citenamefont {Grelier}\ \emph {et~al.}(2022)\citenamefont
		{Grelier}, \citenamefont {Godel}, \citenamefont {Vecchiola}, \citenamefont
		{Collin}, \citenamefont {Bouzehouane}, \citenamefont {Fert}, \citenamefont
		{Cros},\ and\ \citenamefont {Reyren}}]{Grelier2022}%
	\BibitemOpen
	\bibfield  {author} {\bibinfo {author} {\bibfnamefont {M.}~\bibnamefont
			{Grelier}}, \bibinfo {author} {\bibfnamefont {F.}~\bibnamefont {Godel}},
		\bibinfo {author} {\bibfnamefont {A.}~\bibnamefont {Vecchiola}}, \bibinfo
		{author} {\bibfnamefont {S.}~\bibnamefont {Collin}}, \bibinfo {author}
		{\bibfnamefont {K.}~\bibnamefont {Bouzehouane}}, \bibinfo {author}
		{\bibfnamefont {A.}~\bibnamefont {Fert}}, \bibinfo {author} {\bibfnamefont
			{V.}~\bibnamefont {Cros}}, \ and\ \bibinfo {author} {\bibfnamefont
			{N.}~\bibnamefont {Reyren}},\ }\bibfield  {title} {\enquote {\bibinfo {title}
			{Three-dimensional skyrmionic cocoons in magnetic multilayers},}\ }\href
	{\doibase 10.1038/s41467-022-34370-x} {\bibfield  {journal} {\bibinfo
			{journal} {Nature Communications}\ }\textbf {\bibinfo {volume} {13}},\
		\bibinfo {pages} {6843} (\bibinfo {year} {2022})}\BibitemShut {NoStop}%
	\bibitem [{\citenamefont {Krishnia}\ \emph {et~al.}(2023)\citenamefont
		{Krishnia}, \citenamefont {Sassi}, \citenamefont {Ajejas}, \citenamefont
		{Sebe}, \citenamefont {Reyren}, \citenamefont {Collin}, \citenamefont
		{Denneulin}, \citenamefont {Kov{\'a}cs}, \citenamefont {Dunin-Borkowski},
		\citenamefont {Fert}, \citenamefont {George}, \citenamefont {Cros},\ and\
		\citenamefont {Jaffr{\`e}s}}]{krishnia2022}%
	\BibitemOpen
	\bibfield  {author} {\bibinfo {author} {\bibfnamefont {S.}~\bibnamefont
			{Krishnia}}, \bibinfo {author} {\bibfnamefont {Y.}~\bibnamefont {Sassi}},
		\bibinfo {author} {\bibfnamefont {F.}~\bibnamefont {Ajejas}}, \bibinfo
		{author} {\bibfnamefont {N.}~\bibnamefont {Sebe}}, \bibinfo {author}
		{\bibfnamefont {N.}~\bibnamefont {Reyren}}, \bibinfo {author} {\bibfnamefont
			{S.}~\bibnamefont {Collin}}, \bibinfo {author} {\bibfnamefont
			{T.}~\bibnamefont {Denneulin}}, \bibinfo {author} {\bibfnamefont
			{A.}~\bibnamefont {Kov{\'a}cs}}, \bibinfo {author} {\bibfnamefont {R.~E.}\
			\bibnamefont {Dunin-Borkowski}}, \bibinfo {author} {\bibfnamefont
			{A.}~\bibnamefont {Fert}}, \bibinfo {author} {\bibfnamefont {J.-M.}\
			\bibnamefont {George}}, \bibinfo {author} {\bibfnamefont {V.}~\bibnamefont
			{Cros}}, \ and\ \bibinfo {author} {\bibfnamefont {H.}~\bibnamefont
			{Jaffr{\`e}s}},\ }\bibfield  {title} {\enquote {\bibinfo {title} {{Large
					interfacial Rashba interaction generating strong spin--orbit torques in
					atomically thin metallic heterostructures}},}\ }\href {\doibase
		10.1021/acs.nanolett.2c05091} {\bibfield  {journal} {\bibinfo  {journal}
			{Nano Letters}\ }\textbf {\bibinfo {volume} {23}},\ \bibinfo {pages}
		{6785--6791} (\bibinfo {year} {2023})}\BibitemShut {NoStop}%
	\bibitem [{\citenamefont {Jiang}\ \emph {et~al.}(2017)\citenamefont {Jiang},
		\citenamefont {Zhang}, \citenamefont {Yu}, \citenamefont {Zhang},
		\citenamefont {Wang}, \citenamefont {Benjamin~Jungfleisch}, \citenamefont
		{Pearson}, \citenamefont {Cheng}, \citenamefont {Heinonen}, \citenamefont
		{Wang}, \citenamefont {Zhou}, \citenamefont {Hoffmann},\ and\ \citenamefont
		{te~Velthuis}}]{Jiang2017}%
	\BibitemOpen
	\bibfield  {author} {\bibinfo {author} {\bibfnamefont {W.}~\bibnamefont
			{Jiang}}, \bibinfo {author} {\bibfnamefont {X.}~\bibnamefont {Zhang}},
		\bibinfo {author} {\bibfnamefont {G.}~\bibnamefont {Yu}}, \bibinfo {author}
		{\bibfnamefont {W.}~\bibnamefont {Zhang}}, \bibinfo {author} {\bibfnamefont
			{X.}~\bibnamefont {Wang}}, \bibinfo {author} {\bibfnamefont {M.}~\bibnamefont
			{Benjamin~Jungfleisch}}, \bibinfo {author} {\bibfnamefont {J.~E.}\
			\bibnamefont {Pearson}}, \bibinfo {author} {\bibfnamefont {X.}~\bibnamefont
			{Cheng}}, \bibinfo {author} {\bibfnamefont {O.}~\bibnamefont {Heinonen}},
		\bibinfo {author} {\bibfnamefont {K.~L.}\ \bibnamefont {Wang}}, \bibinfo
		{author} {\bibfnamefont {Y.}~\bibnamefont {Zhou}}, \bibinfo {author}
		{\bibfnamefont {A.}~\bibnamefont {Hoffmann}}, \ and\ \bibinfo {author}
		{\bibfnamefont {S.~G.~E.}\ \bibnamefont {te~Velthuis}},\ }\bibfield  {title}
	{\enquote {\bibinfo {title} {Direct observation of the skyrmion hall
				effect},}\ }\href {\doibase 10.1038/nphys3883} {\bibfield  {journal}
		{\bibinfo  {journal} {Nature Physics}\ }\textbf {\bibinfo {volume} {13}},\
		\bibinfo {pages} {162--169} (\bibinfo {year} {2017})}\BibitemShut {NoStop}%
	\bibitem [{\citenamefont {Fert}, \citenamefont {Cros},\ and\ \citenamefont
		{Sampaio}(2013)}]{Fert2013}%
	\BibitemOpen
	\bibfield  {author} {\bibinfo {author} {\bibfnamefont {A.}~\bibnamefont
			{Fert}}, \bibinfo {author} {\bibfnamefont {V.}~\bibnamefont {Cros}}, \ and\
		\bibinfo {author} {\bibfnamefont {J.}~\bibnamefont {Sampaio}},\ }\bibfield
	{title} {\enquote {\bibinfo {title} {Skyrmions on the track},}\ }\href
	{\doibase 10.1038/nnano.2013.29} {\bibfield  {journal} {\bibinfo  {journal}
			{Nature Nanotechnology}\ }\textbf {\bibinfo {volume} {8}},\ \bibinfo {pages}
		{152--156} (\bibinfo {year} {2013})}\BibitemShut {NoStop}%
	\bibitem [{\citenamefont {He}\ \emph {et~al.}(2023)\citenamefont {He},
		\citenamefont {Tomasello}, \citenamefont {Luo}, \citenamefont {Zhang},
		\citenamefont {Nie}, \citenamefont {Carpentieri}, \citenamefont {Han},
		\citenamefont {Finocchio},\ and\ \citenamefont {Yu}}]{He2023}%
	\BibitemOpen
	\bibfield  {author} {\bibinfo {author} {\bibfnamefont {B.}~\bibnamefont
			{He}}, \bibinfo {author} {\bibfnamefont {R.}~\bibnamefont {Tomasello}},
		\bibinfo {author} {\bibfnamefont {X.}~\bibnamefont {Luo}}, \bibinfo {author}
		{\bibfnamefont {R.}~\bibnamefont {Zhang}}, \bibinfo {author} {\bibfnamefont
			{Z.}~\bibnamefont {Nie}}, \bibinfo {author} {\bibfnamefont {M.}~\bibnamefont
			{Carpentieri}}, \bibinfo {author} {\bibfnamefont {X.}~\bibnamefont {Han}},
		\bibinfo {author} {\bibfnamefont {G.}~\bibnamefont {Finocchio}}, \ and\
		\bibinfo {author} {\bibfnamefont {G.}~\bibnamefont {Yu}},\ }\bibfield
	{title} {\enquote {\bibinfo {title} {All-electrical 9-bit skyrmion-based
				racetrack memory designed with laser irradiation},}\ }\bibfield  {booktitle}
	{\emph {\bibinfo {booktitle} {Nano Letters}},\ }\href {\doibase
		10.1021/acs.nanolett.3c02978} {\bibfield  {journal} {\bibinfo  {journal}
			{Nano Letters}\ }\textbf {\bibinfo {volume} {23}},\ \bibinfo {pages}
		{9482--9490} (\bibinfo {year} {2023})}\BibitemShut {NoStop}%
	\bibitem [{\citenamefont {Figueiredo-Prestes}\ \emph
		{et~al.}(2021)\citenamefont {Figueiredo-Prestes}, \citenamefont {Krishnia},
		\citenamefont {Collin}, \citenamefont {Roussigné}, \citenamefont
		{Belmeguenai}, \citenamefont {Chérif}, \citenamefont {Zarpellon},
		\citenamefont {Mosca}, \citenamefont {Jaffrès}, \citenamefont {Vila},
		\citenamefont {Reyren},\ and\ \citenamefont
		{George}}]{FigueiredoPrestes2021}%
	\BibitemOpen
	\bibfield  {author} {\bibinfo {author} {\bibfnamefont {N.}~\bibnamefont
			{Figueiredo-Prestes}}, \bibinfo {author} {\bibfnamefont {S.}~\bibnamefont
			{Krishnia}}, \bibinfo {author} {\bibfnamefont {S.}~\bibnamefont {Collin}},
		\bibinfo {author} {\bibfnamefont {Y.}~\bibnamefont {Roussigné}}, \bibinfo
		{author} {\bibfnamefont {M.}~\bibnamefont {Belmeguenai}}, \bibinfo {author}
		{\bibfnamefont {S.~M.}\ \bibnamefont {Chérif}}, \bibinfo {author}
		{\bibfnamefont {J.}~\bibnamefont {Zarpellon}}, \bibinfo {author}
		{\bibfnamefont {D.~H.}\ \bibnamefont {Mosca}}, \bibinfo {author}
		{\bibfnamefont {H.}~\bibnamefont {Jaffrès}}, \bibinfo {author}
		{\bibfnamefont {L.}~\bibnamefont {Vila}}, \bibinfo {author} {\bibfnamefont
			{N.}~\bibnamefont {Reyren}}, \ and\ \bibinfo {author} {\bibfnamefont {J.-M.}\
			\bibnamefont {George}},\ }\bibfield  {title} {\enquote {\bibinfo {title}
			{{Magnetization switching and deterministic nucleation in Co/Ni multilayered
					disks induced by spin–orbit torques}},}\ }\href {\doibase
		10.1063/5.0050641} {\bibfield  {journal} {\bibinfo  {journal} {Applied
				Physics Letters}\ }\textbf {\bibinfo {volume} {119}},\ \bibinfo {pages}
		{032410} (\bibinfo {year} {2021})}\BibitemShut {NoStop}%
	\bibitem [{\citenamefont {Ambrogio}\ \emph {et~al.}(2023)\citenamefont
		{Ambrogio}, \citenamefont {Narayanan}, \citenamefont {Okazaki}, \citenamefont
		{Fasoli}, \citenamefont {Mackin}, \citenamefont {Hosokawa}, \citenamefont
		{Nomura}, \citenamefont {Yasuda}, \citenamefont {Chen}, \citenamefont {Friz},
		\citenamefont {Ishii}, \citenamefont {Luquin}, \citenamefont {Kohda},
		\citenamefont {Saulnier}, \citenamefont {Brew}, \citenamefont {Choi},
		\citenamefont {Ok}, \citenamefont {Philip}, \citenamefont {Chan},
		\citenamefont {Silvestre}, \citenamefont {Ahsan}, \citenamefont {Narayanan},
		\citenamefont {Tsai},\ and\ \citenamefont {Burr}}]{Ambrogio2023}%
	\BibitemOpen
	\bibfield  {author} {\bibinfo {author} {\bibfnamefont {S.}~\bibnamefont
			{Ambrogio}}, \bibinfo {author} {\bibfnamefont {P.}~\bibnamefont {Narayanan}},
		\bibinfo {author} {\bibfnamefont {A.}~\bibnamefont {Okazaki}}, \bibinfo
		{author} {\bibfnamefont {A.}~\bibnamefont {Fasoli}}, \bibinfo {author}
		{\bibfnamefont {C.}~\bibnamefont {Mackin}}, \bibinfo {author} {\bibfnamefont
			{K.}~\bibnamefont {Hosokawa}}, \bibinfo {author} {\bibfnamefont
			{A.}~\bibnamefont {Nomura}}, \bibinfo {author} {\bibfnamefont
			{T.}~\bibnamefont {Yasuda}}, \bibinfo {author} {\bibfnamefont
			{A.}~\bibnamefont {Chen}}, \bibinfo {author} {\bibfnamefont {A.}~\bibnamefont
			{Friz}}, \bibinfo {author} {\bibfnamefont {M.}~\bibnamefont {Ishii}},
		\bibinfo {author} {\bibfnamefont {J.}~\bibnamefont {Luquin}}, \bibinfo
		{author} {\bibfnamefont {Y.}~\bibnamefont {Kohda}}, \bibinfo {author}
		{\bibfnamefont {N.}~\bibnamefont {Saulnier}}, \bibinfo {author}
		{\bibfnamefont {K.}~\bibnamefont {Brew}}, \bibinfo {author} {\bibfnamefont
			{S.}~\bibnamefont {Choi}}, \bibinfo {author} {\bibfnamefont {I.}~\bibnamefont
			{Ok}}, \bibinfo {author} {\bibfnamefont {T.}~\bibnamefont {Philip}}, \bibinfo
		{author} {\bibfnamefont {V.}~\bibnamefont {Chan}}, \bibinfo {author}
		{\bibfnamefont {C.}~\bibnamefont {Silvestre}}, \bibinfo {author}
		{\bibfnamefont {I.}~\bibnamefont {Ahsan}}, \bibinfo {author} {\bibfnamefont
			{V.}~\bibnamefont {Narayanan}}, \bibinfo {author} {\bibfnamefont
			{H.}~\bibnamefont {Tsai}}, \ and\ \bibinfo {author} {\bibfnamefont {G.~W.}\
			\bibnamefont {Burr}},\ }\bibfield  {title} {\enquote {\bibinfo {title} {An
				analog-ai chip for energy-efficient speech recognition and transcription},}\
	}\href {\doibase 10.1038/s41586-023-06337-5} {\bibfield  {journal} {\bibinfo
			{journal} {Nature}\ }\textbf {\bibinfo {volume} {620}},\ \bibinfo {pages}
		{768--775} (\bibinfo {year} {2023})}\BibitemShut {NoStop}%
	\bibitem [{\citenamefont {Torrejon}\ \emph {et~al.}(2017)\citenamefont
		{Torrejon}, \citenamefont {Riou}, \citenamefont {Araujo}, \citenamefont
		{Tsunegi}, \citenamefont {Khalsa}, \citenamefont {Querlioz}, \citenamefont
		{Bortolotti}, \citenamefont {Cros}, \citenamefont {Yakushiji}, \citenamefont
		{Fukushima}, \citenamefont {Kubota}, \citenamefont {Yuasa}, \citenamefont
		{Stiles},\ and\ \citenamefont {Grollier}}]{Torrejon2017}%
	\BibitemOpen
	\bibfield  {author} {\bibinfo {author} {\bibfnamefont {J.}~\bibnamefont
			{Torrejon}}, \bibinfo {author} {\bibfnamefont {M.}~\bibnamefont {Riou}},
		\bibinfo {author} {\bibfnamefont {F.~A.}\ \bibnamefont {Araujo}}, \bibinfo
		{author} {\bibfnamefont {S.}~\bibnamefont {Tsunegi}}, \bibinfo {author}
		{\bibfnamefont {G.}~\bibnamefont {Khalsa}}, \bibinfo {author} {\bibfnamefont
			{D.}~\bibnamefont {Querlioz}}, \bibinfo {author} {\bibfnamefont
			{P.}~\bibnamefont {Bortolotti}}, \bibinfo {author} {\bibfnamefont
			{V.}~\bibnamefont {Cros}}, \bibinfo {author} {\bibfnamefont {K.}~\bibnamefont
			{Yakushiji}}, \bibinfo {author} {\bibfnamefont {A.}~\bibnamefont
			{Fukushima}}, \bibinfo {author} {\bibfnamefont {H.}~\bibnamefont {Kubota}},
		\bibinfo {author} {\bibfnamefont {S.}~\bibnamefont {Yuasa}}, \bibinfo
		{author} {\bibfnamefont {M.~D.}\ \bibnamefont {Stiles}}, \ and\ \bibinfo
		{author} {\bibfnamefont {J.}~\bibnamefont {Grollier}},\ }\bibfield  {title}
	{\enquote {\bibinfo {title} {Neuromorphic computing with nanoscale spintronic
				oscillators},}\ }\href {\doibase 10.1038/nature23011} {\bibfield  {journal}
		{\bibinfo  {journal} {Nature}\ }\textbf {\bibinfo {volume} {547}},\ \bibinfo
		{pages} {428--431} (\bibinfo {year} {2017})}\BibitemShut {NoStop}%
	\bibitem [{\citenamefont {Azghadi}\ \emph {et~al.}(2020)\citenamefont
		{Azghadi}, \citenamefont {Lammie}, \citenamefont {Eshraghian}, \citenamefont
		{Payvand}, \citenamefont {Donati}, \citenamefont {Linares-Barranco},\ and\
		\citenamefont {Indiveri}}]{Azghadi2020}%
	\BibitemOpen
	\bibfield  {author} {\bibinfo {author} {\bibfnamefont {M.~R.}\ \bibnamefont
			{Azghadi}}, \bibinfo {author} {\bibfnamefont {C.}~\bibnamefont {Lammie}},
		\bibinfo {author} {\bibfnamefont {J.~K.}\ \bibnamefont {Eshraghian}},
		\bibinfo {author} {\bibfnamefont {M.}~\bibnamefont {Payvand}}, \bibinfo
		{author} {\bibfnamefont {E.}~\bibnamefont {Donati}}, \bibinfo {author}
		{\bibfnamefont {B.}~\bibnamefont {Linares-Barranco}}, \ and\ \bibinfo
		{author} {\bibfnamefont {G.}~\bibnamefont {Indiveri}},\ }\bibfield  {title}
	{\enquote {\bibinfo {title} {Hardware implementation of deep network
				accelerators towards healthcare and biomedical applications},}\ }\href
	{\doibase 10.1109/TBCAS.2020.3036081} {\bibfield  {journal} {\bibinfo
			{journal} {IEEE Transactions on Biomedical Circuits and Systems}\ }\textbf
		{\bibinfo {volume} {14}},\ \bibinfo {pages} {1138--1159} (\bibinfo {year}
		{2020})}\BibitemShut {NoStop}%
	\bibitem [{\citenamefont {Krishnia}\ \emph {et~al.}(2024)\citenamefont
		{Krishnia}, \citenamefont {Bony}, \citenamefont {Rongione}, \citenamefont
		{Vicente-Arche}, \citenamefont {Denneulin}, \citenamefont {Pezo},
		\citenamefont {Lu}, \citenamefont {Dunin-Borkowski}, \citenamefont {Collin},
		\citenamefont {Fert}, \citenamefont {George}, \citenamefont {Reyren},
		\citenamefont {Cros},\ and\ \citenamefont
		{Jaffrès}}]{krishnia2023quantifying}%
	\BibitemOpen
	\bibfield  {author} {\bibinfo {author} {\bibfnamefont {S.}~\bibnamefont
			{Krishnia}}, \bibinfo {author} {\bibfnamefont {B.}~\bibnamefont {Bony}},
		\bibinfo {author} {\bibfnamefont {E.}~\bibnamefont {Rongione}}, \bibinfo
		{author} {\bibfnamefont {L.~M.}\ \bibnamefont {Vicente-Arche}}, \bibinfo
		{author} {\bibfnamefont {T.}~\bibnamefont {Denneulin}}, \bibinfo {author}
		{\bibfnamefont {A.}~\bibnamefont {Pezo}}, \bibinfo {author} {\bibfnamefont
			{Y.}~\bibnamefont {Lu}}, \bibinfo {author} {\bibfnamefont {R.~E.}\
			\bibnamefont {Dunin-Borkowski}}, \bibinfo {author} {\bibfnamefont
			{S.}~\bibnamefont {Collin}}, \bibinfo {author} {\bibfnamefont
			{A.}~\bibnamefont {Fert}}, \bibinfo {author} {\bibfnamefont {J.-M.}\
			\bibnamefont {George}}, \bibinfo {author} {\bibfnamefont {N.}~\bibnamefont
			{Reyren}}, \bibinfo {author} {\bibfnamefont {V.}~\bibnamefont {Cros}}, \ and\
		\bibinfo {author} {\bibfnamefont {H.}~\bibnamefont {Jaffrès}},\ }\bibfield
	{title} {\enquote {\bibinfo {title} {{Quantifying the large contribution from
					orbital Rashba–Edelstein effect to the effective damping-like torque on
					magnetization}},}\ }\href {\doibase 10.1063/5.0198970} {\bibfield  {journal}
		{\bibinfo  {journal} {APL Materials}\ }\textbf {\bibinfo {volume} {12}},\
		\bibinfo {pages} {051105} (\bibinfo {year} {2024})}\BibitemShut {NoStop}%
	\bibitem [{\citenamefont {Schott}\ \emph {et~al.}(2017)\citenamefont {Schott},
		\citenamefont {Bernand-Mantel}, \citenamefont {Ranno}, \citenamefont
		{Pizzini}, \citenamefont {Vogel}, \citenamefont {B{\'e}a}, \citenamefont
		{Baraduc}, \citenamefont {Auffret}, \citenamefont {Gaudin},\ and\
		\citenamefont {Givord}}]{Schott2017}%
	\BibitemOpen
	\bibfield  {author} {\bibinfo {author} {\bibfnamefont {M.}~\bibnamefont
			{Schott}}, \bibinfo {author} {\bibfnamefont {A.}~\bibnamefont
			{Bernand-Mantel}}, \bibinfo {author} {\bibfnamefont {L.}~\bibnamefont
			{Ranno}}, \bibinfo {author} {\bibfnamefont {S.}~\bibnamefont {Pizzini}},
		\bibinfo {author} {\bibfnamefont {J.}~\bibnamefont {Vogel}}, \bibinfo
		{author} {\bibfnamefont {H.}~\bibnamefont {B{\'e}a}}, \bibinfo {author}
		{\bibfnamefont {C.}~\bibnamefont {Baraduc}}, \bibinfo {author} {\bibfnamefont
			{S.}~\bibnamefont {Auffret}}, \bibinfo {author} {\bibfnamefont
			{G.}~\bibnamefont {Gaudin}}, \ and\ \bibinfo {author} {\bibfnamefont
			{D.}~\bibnamefont {Givord}},\ }\bibfield  {title} {\enquote {\bibinfo {title}
			{The skyrmion switch: Turning magnetic skyrmion bubbles on and off with an
				electric field},}\ }\href {\doibase 10.1021/acs.nanolett.7b00328} {\bibfield
		{journal} {\bibinfo  {journal} {Nano Letters}\ }\textbf {\bibinfo {volume}
			{17}},\ \bibinfo {pages} {3006--3012} (\bibinfo {year} {2017})}\BibitemShut
	{NoStop}%
	\bibitem [{\citenamefont {Bhattacharya}\ \emph {et~al.}(2020)\citenamefont
		{Bhattacharya}, \citenamefont {Razavi}, \citenamefont {Wu}, \citenamefont
		{Dai}, \citenamefont {Wang},\ and\ \citenamefont
		{Atulasimha}}]{Bhattacharya2020}%
	\BibitemOpen
	\bibfield  {author} {\bibinfo {author} {\bibfnamefont {D.}~\bibnamefont
			{Bhattacharya}}, \bibinfo {author} {\bibfnamefont {S.~A.}\ \bibnamefont
			{Razavi}}, \bibinfo {author} {\bibfnamefont {H.}~\bibnamefont {Wu}}, \bibinfo
		{author} {\bibfnamefont {B.}~\bibnamefont {Dai}}, \bibinfo {author}
		{\bibfnamefont {K.~L.}\ \bibnamefont {Wang}}, \ and\ \bibinfo {author}
		{\bibfnamefont {J.}~\bibnamefont {Atulasimha}},\ }\bibfield  {title}
	{\enquote {\bibinfo {title} {Creation and annihilation of non-volatile fixed
				magnetic skyrmions using voltage control of magnetic anisotropy},}\ }\href
	{\doibase 10.1038/s41928-020-0432-x} {\bibfield  {journal} {\bibinfo
			{journal} {Nature Electronics}\ }\textbf {\bibinfo {volume} {3}},\ \bibinfo
		{pages} {539--545} (\bibinfo {year} {2020})}\BibitemShut {NoStop}%
	\bibitem [{\citenamefont {Bernand-Mantel}\ \emph {et~al.}(2013)\citenamefont
		{Bernand-Mantel}, \citenamefont {Herrera-Diez}, \citenamefont {Ranno},
		\citenamefont {Pizzini}, \citenamefont {Vogel}, \citenamefont {Givord},
		\citenamefont {Auffret}, \citenamefont {Boulle}, \citenamefont {Miron},\ and\
		\citenamefont {Gaudin}}]{Bernand-Mantel2013}%
	\BibitemOpen
	\bibfield  {author} {\bibinfo {author} {\bibfnamefont {A.}~\bibnamefont
			{Bernand-Mantel}}, \bibinfo {author} {\bibfnamefont {L.}~\bibnamefont
			{Herrera-Diez}}, \bibinfo {author} {\bibfnamefont {L.}~\bibnamefont {Ranno}},
		\bibinfo {author} {\bibfnamefont {S.}~\bibnamefont {Pizzini}}, \bibinfo
		{author} {\bibfnamefont {J.}~\bibnamefont {Vogel}}, \bibinfo {author}
		{\bibfnamefont {D.}~\bibnamefont {Givord}}, \bibinfo {author} {\bibfnamefont
			{S.}~\bibnamefont {Auffret}}, \bibinfo {author} {\bibfnamefont
			{O.}~\bibnamefont {Boulle}}, \bibinfo {author} {\bibfnamefont {I.~M.}\
			\bibnamefont {Miron}}, \ and\ \bibinfo {author} {\bibfnamefont
			{G.}~\bibnamefont {Gaudin}},\ }\bibfield  {title} {\enquote {\bibinfo {title}
			{Electric-field control of domain wall nucleation and pinning in a metallic
				ferromagnet},}\ }\href {\doibase 10.1063/1.4798506} {\bibfield  {journal}
		{\bibinfo  {journal} {Applied Physics Letters}\ }\textbf {\bibinfo {volume}
			{102}},\ \bibinfo {pages} {122406} (\bibinfo {year} {2013})}\BibitemShut
	{NoStop}%
	\bibitem [{\citenamefont {Bauer}\ \emph {et~al.}(2015)\citenamefont {Bauer},
		\citenamefont {Yao}, \citenamefont {Tan}, \citenamefont {Agrawal},
		\citenamefont {Emori}, \citenamefont {Tuller}, \citenamefont {van Dijken},\
		and\ \citenamefont {Beach}}]{Bauer2015}%
	\BibitemOpen
	\bibfield  {author} {\bibinfo {author} {\bibfnamefont {U.}~\bibnamefont
			{Bauer}}, \bibinfo {author} {\bibfnamefont {L.}~\bibnamefont {Yao}}, \bibinfo
		{author} {\bibfnamefont {A.~J.}\ \bibnamefont {Tan}}, \bibinfo {author}
		{\bibfnamefont {P.}~\bibnamefont {Agrawal}}, \bibinfo {author} {\bibfnamefont
			{S.}~\bibnamefont {Emori}}, \bibinfo {author} {\bibfnamefont {H.~L.}\
			\bibnamefont {Tuller}}, \bibinfo {author} {\bibfnamefont {S.}~\bibnamefont
			{van Dijken}}, \ and\ \bibinfo {author} {\bibfnamefont {G.~S.~D.}\
			\bibnamefont {Beach}},\ }\bibfield  {title} {\enquote {\bibinfo {title}
			{Magneto-ionic control of interfacial magnetism},}\ }\href {\doibase
		10.1038/nmat4134} {\bibfield  {journal} {\bibinfo  {journal} {Nature
				Materials}\ }\textbf {\bibinfo {volume} {14}},\ \bibinfo {pages} {174--181}
		(\bibinfo {year} {2015})}\BibitemShut {NoStop}%
	\bibitem [{\citenamefont {Herrera~Diez}\ \emph {et~al.}(2019)\citenamefont
		{Herrera~Diez}, \citenamefont {Liu}, \citenamefont {Gilbert}, \citenamefont
		{Belmeguenai}, \citenamefont {Vogel}, \citenamefont {Pizzini}, \citenamefont
		{Martinez}, \citenamefont {Lamperti}, \citenamefont {Mohammedi},
		\citenamefont {Laborieux}, \citenamefont {Roussign\'e}, \citenamefont
		{Grutter}, \citenamefont {Arenholtz}, \citenamefont {Quarterman},
		\citenamefont {Maranville}, \citenamefont {Ono}, \citenamefont {Hadri},
		\citenamefont {Tolley}, \citenamefont {Fullerton}, \citenamefont
		{Sanchez-Tejerina}, \citenamefont {Stashkevich}, \citenamefont {Ch\'erif},
		\citenamefont {Kent}, \citenamefont {Querlioz}, \citenamefont {Langer},
		\citenamefont {Ocker},\ and\ \citenamefont {Ravelosona}}]{HerreraDiez2019}%
	\BibitemOpen
	\bibfield  {author} {\bibinfo {author} {\bibfnamefont {L.}~\bibnamefont
			{Herrera~Diez}}, \bibinfo {author} {\bibfnamefont {Y.}~\bibnamefont {Liu}},
		\bibinfo {author} {\bibfnamefont {D.}~\bibnamefont {Gilbert}}, \bibinfo
		{author} {\bibfnamefont {M.}~\bibnamefont {Belmeguenai}}, \bibinfo {author}
		{\bibfnamefont {J.}~\bibnamefont {Vogel}}, \bibinfo {author} {\bibfnamefont
			{S.}~\bibnamefont {Pizzini}}, \bibinfo {author} {\bibfnamefont
			{E.}~\bibnamefont {Martinez}}, \bibinfo {author} {\bibfnamefont
			{A.}~\bibnamefont {Lamperti}}, \bibinfo {author} {\bibfnamefont
			{J.}~\bibnamefont {Mohammedi}}, \bibinfo {author} {\bibfnamefont
			{A.}~\bibnamefont {Laborieux}}, \bibinfo {author} {\bibfnamefont
			{Y.}~\bibnamefont {Roussign\'e}}, \bibinfo {author} {\bibfnamefont
			{A.}~\bibnamefont {Grutter}}, \bibinfo {author} {\bibfnamefont
			{E.}~\bibnamefont {Arenholtz}}, \bibinfo {author} {\bibfnamefont
			{P.}~\bibnamefont {Quarterman}}, \bibinfo {author} {\bibfnamefont
			{B.}~\bibnamefont {Maranville}}, \bibinfo {author} {\bibfnamefont
			{S.}~\bibnamefont {Ono}}, \bibinfo {author} {\bibfnamefont {M.~S.~E.}\
			\bibnamefont {Hadri}}, \bibinfo {author} {\bibfnamefont {R.}~\bibnamefont
			{Tolley}}, \bibinfo {author} {\bibfnamefont {E.}~\bibnamefont {Fullerton}},
		\bibinfo {author} {\bibfnamefont {L.}~\bibnamefont {Sanchez-Tejerina}},
		\bibinfo {author} {\bibfnamefont {A.}~\bibnamefont {Stashkevich}}, \bibinfo
		{author} {\bibfnamefont {S.}~\bibnamefont {Ch\'erif}}, \bibinfo {author}
		{\bibfnamefont {A.}~\bibnamefont {Kent}}, \bibinfo {author} {\bibfnamefont
			{D.}~\bibnamefont {Querlioz}}, \bibinfo {author} {\bibfnamefont
			{J.}~\bibnamefont {Langer}}, \bibinfo {author} {\bibfnamefont
			{B.}~\bibnamefont {Ocker}}, \ and\ \bibinfo {author} {\bibfnamefont
			{D.}~\bibnamefont {Ravelosona}},\ }\bibfield  {title} {\enquote {\bibinfo
			{title} {Nonvolatile ionic modification of the dzyaloshinskii-moriya
				interaction},}\ }\href {\doibase 10.1103/PhysRevApplied.12.034005} {\bibfield
		{journal} {\bibinfo  {journal} {Phys. Rev. Appl.}\ }\textbf {\bibinfo
			{volume} {12}},\ \bibinfo {pages} {034005} (\bibinfo {year}
		{2019})}\BibitemShut {NoStop}%
	\bibitem [{\citenamefont {Srivastava}\ \emph {et~al.}(2018)\citenamefont
		{Srivastava}, \citenamefont {Schott}, \citenamefont {Juge}, \citenamefont
		{K{\v r}i{\v z}{\'a}kov{\'a}}, \citenamefont {Belmeguenai}, \citenamefont
		{Roussign{\'e}}, \citenamefont {Bernand-Mantel}, \citenamefont {Ranno},
		\citenamefont {Pizzini}, \citenamefont {Ch{\'e}rif}, \citenamefont
		{Stashkevich}, \citenamefont {Auffret}, \citenamefont {Boulle}, \citenamefont
		{Gaudin}, \citenamefont {Chshiev}, \citenamefont {Baraduc},\ and\
		\citenamefont {B{\'e}a}}]{Srivastava2018}%
	\BibitemOpen
	\bibfield  {author} {\bibinfo {author} {\bibfnamefont {T.}~\bibnamefont
			{Srivastava}}, \bibinfo {author} {\bibfnamefont {M.}~\bibnamefont {Schott}},
		\bibinfo {author} {\bibfnamefont {R.}~\bibnamefont {Juge}}, \bibinfo {author}
		{\bibfnamefont {V.}~\bibnamefont {K{\v r}i{\v z}{\'a}kov{\'a}}}, \bibinfo
		{author} {\bibfnamefont {M.}~\bibnamefont {Belmeguenai}}, \bibinfo {author}
		{\bibfnamefont {Y.}~\bibnamefont {Roussign{\'e}}}, \bibinfo {author}
		{\bibfnamefont {A.}~\bibnamefont {Bernand-Mantel}}, \bibinfo {author}
		{\bibfnamefont {L.}~\bibnamefont {Ranno}}, \bibinfo {author} {\bibfnamefont
			{S.}~\bibnamefont {Pizzini}}, \bibinfo {author} {\bibfnamefont {S.-M.}\
			\bibnamefont {Ch{\'e}rif}}, \bibinfo {author} {\bibfnamefont
			{A.}~\bibnamefont {Stashkevich}}, \bibinfo {author} {\bibfnamefont
			{S.}~\bibnamefont {Auffret}}, \bibinfo {author} {\bibfnamefont
			{O.}~\bibnamefont {Boulle}}, \bibinfo {author} {\bibfnamefont
			{G.}~\bibnamefont {Gaudin}}, \bibinfo {author} {\bibfnamefont
			{M.}~\bibnamefont {Chshiev}}, \bibinfo {author} {\bibfnamefont
			{C.}~\bibnamefont {Baraduc}}, \ and\ \bibinfo {author} {\bibfnamefont
			{H.}~\bibnamefont {B{\'e}a}},\ }\bibfield  {title} {\enquote {\bibinfo
			{title} {Large-voltage tuning of dzyaloshinskii--moriya interactions: A route
				toward dynamic control of skyrmion chirality},}\ }\href {\doibase
		10.1021/acs.nanolett.8b01502} {\bibfield  {journal} {\bibinfo  {journal}
			{Nano Letters}\ }\textbf {\bibinfo {volume} {18}},\ \bibinfo {pages}
		{4871--4877} (\bibinfo {year} {2018})}\BibitemShut {NoStop}%
	\bibitem [{\citenamefont {Fillion}\ \emph {et~al.}(2022)\citenamefont
		{Fillion}, \citenamefont {Fischer}, \citenamefont {Kumar}, \citenamefont
		{Fassatoui}, \citenamefont {Pizzini}, \citenamefont {Ranno}, \citenamefont
		{Ourdani}, \citenamefont {Belmeguenai}, \citenamefont {Roussign{\'e}},
		\citenamefont {Ch{\'e}rif}, \citenamefont {Auffret}, \citenamefont {Joumard},
		\citenamefont {Boulle}, \citenamefont {Gaudin}, \citenamefont
		{Buda-Prejbeanu}, \citenamefont {Baraduc},\ and\ \citenamefont
		{B{\'e}a}}]{Fillion2022}%
	\BibitemOpen
	\bibfield  {author} {\bibinfo {author} {\bibfnamefont {C.-E.}\ \bibnamefont
			{Fillion}}, \bibinfo {author} {\bibfnamefont {J.}~\bibnamefont {Fischer}},
		\bibinfo {author} {\bibfnamefont {R.}~\bibnamefont {Kumar}}, \bibinfo
		{author} {\bibfnamefont {A.}~\bibnamefont {Fassatoui}}, \bibinfo {author}
		{\bibfnamefont {S.}~\bibnamefont {Pizzini}}, \bibinfo {author} {\bibfnamefont
			{L.}~\bibnamefont {Ranno}}, \bibinfo {author} {\bibfnamefont
			{D.}~\bibnamefont {Ourdani}}, \bibinfo {author} {\bibfnamefont
			{M.}~\bibnamefont {Belmeguenai}}, \bibinfo {author} {\bibfnamefont
			{Y.}~\bibnamefont {Roussign{\'e}}}, \bibinfo {author} {\bibfnamefont {S.-M.}\
			\bibnamefont {Ch{\'e}rif}}, \bibinfo {author} {\bibfnamefont
			{S.}~\bibnamefont {Auffret}}, \bibinfo {author} {\bibfnamefont
			{I.}~\bibnamefont {Joumard}}, \bibinfo {author} {\bibfnamefont
			{O.}~\bibnamefont {Boulle}}, \bibinfo {author} {\bibfnamefont
			{G.}~\bibnamefont {Gaudin}}, \bibinfo {author} {\bibfnamefont
			{L.}~\bibnamefont {Buda-Prejbeanu}}, \bibinfo {author} {\bibfnamefont
			{C.}~\bibnamefont {Baraduc}}, \ and\ \bibinfo {author} {\bibfnamefont
			{H.}~\bibnamefont {B{\'e}a}},\ }\bibfield  {title} {\enquote {\bibinfo
			{title} {Gate-controlled skyrmion and domain wall chirality},}\ }\href
	{\doibase 10.1038/s41467-022-32959-w} {\bibfield  {journal} {\bibinfo
			{journal} {Nature Communications}\ }\textbf {\bibinfo {volume} {13}},\
		\bibinfo {pages} {5257} (\bibinfo {year} {2022})}\BibitemShut {NoStop}%
	\bibitem [{\citenamefont {Dai}\ \emph {et~al.}(2023)\citenamefont {Dai},
		\citenamefont {Wu}, \citenamefont {Razavi}, \citenamefont {Xu}, \citenamefont
		{He}, \citenamefont {Shu}, \citenamefont {Jackson}, \citenamefont {Mahfouzi},
		\citenamefont {Huang}, \citenamefont {Pan}, \citenamefont {Cheng},
		\citenamefont {Qu}, \citenamefont {Wang}, \citenamefont {Tai}, \citenamefont
		{Wong}, \citenamefont {Kioussis},\ and\ \citenamefont {Wang}}]{Dai2023}%
	\BibitemOpen
	\bibfield  {author} {\bibinfo {author} {\bibfnamefont {B.}~\bibnamefont
			{Dai}}, \bibinfo {author} {\bibfnamefont {D.}~\bibnamefont {Wu}}, \bibinfo
		{author} {\bibfnamefont {S.~A.}\ \bibnamefont {Razavi}}, \bibinfo {author}
		{\bibfnamefont {S.}~\bibnamefont {Xu}}, \bibinfo {author} {\bibfnamefont
			{H.}~\bibnamefont {He}}, \bibinfo {author} {\bibfnamefont {Q.}~\bibnamefont
			{Shu}}, \bibinfo {author} {\bibfnamefont {M.}~\bibnamefont {Jackson}},
		\bibinfo {author} {\bibfnamefont {F.}~\bibnamefont {Mahfouzi}}, \bibinfo
		{author} {\bibfnamefont {H.}~\bibnamefont {Huang}}, \bibinfo {author}
		{\bibfnamefont {Q.}~\bibnamefont {Pan}}, \bibinfo {author} {\bibfnamefont
			{Y.}~\bibnamefont {Cheng}}, \bibinfo {author} {\bibfnamefont
			{T.}~\bibnamefont {Qu}}, \bibinfo {author} {\bibfnamefont {T.}~\bibnamefont
			{Wang}}, \bibinfo {author} {\bibfnamefont {L.}~\bibnamefont {Tai}}, \bibinfo
		{author} {\bibfnamefont {K.}~\bibnamefont {Wong}}, \bibinfo {author}
		{\bibfnamefont {N.}~\bibnamefont {Kioussis}}, \ and\ \bibinfo {author}
		{\bibfnamefont {K.~L.}\ \bibnamefont {Wang}},\ }\bibfield  {title} {\enquote
		{\bibinfo {title} {Electric field manipulation of spin chirality and skyrmion
				dynamic},}\ }\href {\doibase 10.1126/sciadv.ade6836} {\bibfield  {journal}
		{\bibinfo  {journal} {Science Advances}\ }\textbf {\bibinfo {volume} {9}},\
		\bibinfo {pages} {eade6836} (\bibinfo {year} {2023})}\BibitemShut {NoStop}%
	\bibitem [{\citenamefont {Mishra}, \citenamefont {Kumar},\ and\ \citenamefont
		{Yang}(2019)}]{Mishra2019}%
	\BibitemOpen
	\bibfield  {author} {\bibinfo {author} {\bibfnamefont {R.}~\bibnamefont
			{Mishra}}, \bibinfo {author} {\bibfnamefont {D.}~\bibnamefont {Kumar}}, \
		and\ \bibinfo {author} {\bibfnamefont {H.}~\bibnamefont {Yang}},\ }\bibfield
	{title} {\enquote {\bibinfo {title} {Oxygen-migration-based spintronic device
				emulating a biological synapse},}\ }\href {\doibase
		10.1103/PhysRevApplied.11.054065} {\bibfield  {journal} {\bibinfo  {journal}
			{Phys. Rev. Appl.}\ }\textbf {\bibinfo {volume} {11}},\ \bibinfo {pages}
		{054065} (\bibinfo {year} {2019})}\BibitemShut {NoStop}%
	\bibitem [{\citenamefont {da~C\^amara Santa Clara~Gomes}\ \emph
		{et~al.}(2024)\citenamefont {da~C\^amara Santa Clara~Gomes}, \citenamefont
		{Bhatnagar-Sch\"offmann}, \citenamefont {Krishnia}, \citenamefont {Sassi},
		\citenamefont {Sanz-Hern\'andez}, \citenamefont {Reyren}, \citenamefont
		{Martin}, \citenamefont {Brunnett}, \citenamefont {Collin}, \citenamefont
		{Godel}, \citenamefont {Ono}, \citenamefont {Querlioz}, \citenamefont
		{Ravelosona}, \citenamefont {Cros}, \citenamefont {Grollier}, \citenamefont
		{Seneor},\ and\ \citenamefont {Herrera~Diez}}]{Tristan2024}%
	\BibitemOpen
	\bibfield  {author} {\bibinfo {author} {\bibfnamefont {T.}~\bibnamefont
			{da~C\^amara Santa Clara~Gomes}}, \bibinfo {author} {\bibfnamefont
			{T.}~\bibnamefont {Bhatnagar-Sch\"offmann}}, \bibinfo {author} {\bibfnamefont
			{S.}~\bibnamefont {Krishnia}}, \bibinfo {author} {\bibfnamefont
			{Y.}~\bibnamefont {Sassi}}, \bibinfo {author} {\bibfnamefont
			{D.}~\bibnamefont {Sanz-Hern\'andez}}, \bibinfo {author} {\bibfnamefont
			{N.}~\bibnamefont {Reyren}}, \bibinfo {author} {\bibfnamefont {M.-B.}\
			\bibnamefont {Martin}}, \bibinfo {author} {\bibfnamefont {F.}~\bibnamefont
			{Brunnett}}, \bibinfo {author} {\bibfnamefont {S.}~\bibnamefont {Collin}},
		\bibinfo {author} {\bibfnamefont {F.}~\bibnamefont {Godel}}, \bibinfo
		{author} {\bibfnamefont {S.}~\bibnamefont {Ono}}, \bibinfo {author}
		{\bibfnamefont {D.}~\bibnamefont {Querlioz}}, \bibinfo {author}
		{\bibfnamefont {D.}~\bibnamefont {Ravelosona}}, \bibinfo {author}
		{\bibfnamefont {V.}~\bibnamefont {Cros}}, \bibinfo {author} {\bibfnamefont
			{J.}~\bibnamefont {Grollier}}, \bibinfo {author} {\bibfnamefont
			{P.}~\bibnamefont {Seneor}}, \ and\ \bibinfo {author} {\bibfnamefont
			{L.}~\bibnamefont {Herrera~Diez}},\ }\bibfield  {title} {\enquote {\bibinfo
			{title} {Control of the magnetic anisotropy in multirepeat pt/co/al
				heterostructures using magnetoionic gating},}\ }\href {\doibase
		10.1103/PhysRevApplied.21.024010} {\bibfield  {journal} {\bibinfo  {journal}
			{Phys. Rev. Appl.}\ }\textbf {\bibinfo {volume} {21}},\ \bibinfo {pages}
		{024010} (\bibinfo {year} {2024})}\BibitemShut {NoStop}%
	\bibitem [{\citenamefont {Legrand}\ \emph {et~al.}(2020)\citenamefont
		{Legrand}, \citenamefont {Maccariello}, \citenamefont {Ajejas}, \citenamefont
		{Collin}, \citenamefont {Vecchiola}, \citenamefont {Bouzehouane},
		\citenamefont {Reyren}, \citenamefont {Cros},\ and\ \citenamefont
		{Fert}}]{Legrand2019}%
	\BibitemOpen
	\bibfield  {author} {\bibinfo {author} {\bibfnamefont {W.}~\bibnamefont
			{Legrand}}, \bibinfo {author} {\bibfnamefont {D.}~\bibnamefont
			{Maccariello}}, \bibinfo {author} {\bibfnamefont {F.}~\bibnamefont {Ajejas}},
		\bibinfo {author} {\bibfnamefont {S.}~\bibnamefont {Collin}}, \bibinfo
		{author} {\bibfnamefont {A.}~\bibnamefont {Vecchiola}}, \bibinfo {author}
		{\bibfnamefont {K.}~\bibnamefont {Bouzehouane}}, \bibinfo {author}
		{\bibfnamefont {N.}~\bibnamefont {Reyren}}, \bibinfo {author} {\bibfnamefont
			{V.}~\bibnamefont {Cros}}, \ and\ \bibinfo {author} {\bibfnamefont
			{A.}~\bibnamefont {Fert}},\ }\bibfield  {title} {\enquote {\bibinfo {title}
			{Room-temperature stabilization of antiferromagnetic skyrmions in synthetic
				antiferromagnets},}\ }\href {\doibase 10.1038/s41563-019-0468-3} {\bibfield
		{journal} {\bibinfo  {journal} {Nature Materials}\ }\textbf {\bibinfo
			{volume} {19}},\ \bibinfo {pages} {34--42} (\bibinfo {year}
		{2020})}\BibitemShut {NoStop}%
	\bibitem [{\citenamefont {Juge}\ \emph {et~al.}(2019)\citenamefont {Juge},
		\citenamefont {Je}, \citenamefont {Chaves}, \citenamefont {Buda-Prejbeanu},
		\citenamefont {Pe\~na Garcia}, \citenamefont {Nath}, \citenamefont {Miron},
		\citenamefont {Rana}, \citenamefont {Aballe}, \citenamefont {Foerster},
		\citenamefont {Genuzio}, \citenamefont {Mentes}, \citenamefont {Locatelli},
		\citenamefont {Maccherozzi}, \citenamefont {Dhesi}, \citenamefont
		{Belmeguenai}, \citenamefont {Roussign\'e}, \citenamefont {Auffret},
		\citenamefont {Pizzini}, \citenamefont {Gaudin}, \citenamefont {Vogel},\ and\
		\citenamefont {Boulle}}]{Juge2019}%
	\BibitemOpen
	\bibfield  {author} {\bibinfo {author} {\bibfnamefont {R.}~\bibnamefont
			{Juge}}, \bibinfo {author} {\bibfnamefont {S.-G.}\ \bibnamefont {Je}},
		\bibinfo {author} {\bibfnamefont {D.~d.~S.}\ \bibnamefont {Chaves}}, \bibinfo
		{author} {\bibfnamefont {L.~D.}\ \bibnamefont {Buda-Prejbeanu}}, \bibinfo
		{author} {\bibfnamefont {J.}~\bibnamefont {Pe\~na Garcia}}, \bibinfo {author}
		{\bibfnamefont {J.}~\bibnamefont {Nath}}, \bibinfo {author} {\bibfnamefont
			{I.~M.}\ \bibnamefont {Miron}}, \bibinfo {author} {\bibfnamefont {K.~G.}\
			\bibnamefont {Rana}}, \bibinfo {author} {\bibfnamefont {L.}~\bibnamefont
			{Aballe}}, \bibinfo {author} {\bibfnamefont {M.}~\bibnamefont {Foerster}},
		\bibinfo {author} {\bibfnamefont {F.}~\bibnamefont {Genuzio}}, \bibinfo
		{author} {\bibfnamefont {T.~O.}\ \bibnamefont {Mentes}}, \bibinfo {author}
		{\bibfnamefont {A.}~\bibnamefont {Locatelli}}, \bibinfo {author}
		{\bibfnamefont {F.}~\bibnamefont {Maccherozzi}}, \bibinfo {author}
		{\bibfnamefont {S.~S.}\ \bibnamefont {Dhesi}}, \bibinfo {author}
		{\bibfnamefont {M.}~\bibnamefont {Belmeguenai}}, \bibinfo {author}
		{\bibfnamefont {Y.}~\bibnamefont {Roussign\'e}}, \bibinfo {author}
		{\bibfnamefont {S.}~\bibnamefont {Auffret}}, \bibinfo {author} {\bibfnamefont
			{S.}~\bibnamefont {Pizzini}}, \bibinfo {author} {\bibfnamefont
			{G.}~\bibnamefont {Gaudin}}, \bibinfo {author} {\bibfnamefont
			{J.}~\bibnamefont {Vogel}}, \ and\ \bibinfo {author} {\bibfnamefont
			{O.}~\bibnamefont {Boulle}},\ }\bibfield  {title} {\enquote {\bibinfo {title}
			{Current-driven skyrmion dynamics and drive-dependent skyrmion hall effect in
				an ultrathin film},}\ }\href {\doibase 10.1103/PhysRevApplied.12.044007}
	{\bibfield  {journal} {\bibinfo  {journal} {Phys. Rev. Appl.}\ }\textbf
		{\bibinfo {volume} {12}},\ \bibinfo {pages} {044007} (\bibinfo {year}
		{2019})}\BibitemShut {NoStop}%
	\bibitem [{\citenamefont {Litzius}\ \emph {et~al.}(2020)\citenamefont
		{Litzius}, \citenamefont {Leliaert}, \citenamefont {Bassirian}, \citenamefont
		{Rodrigues}, \citenamefont {Kromin}, \citenamefont {Lemesh}, \citenamefont
		{Zazvorka}, \citenamefont {Lee}, \citenamefont {Mulkers}, \citenamefont
		{Kerber}, \citenamefont {Heinze}, \citenamefont {Keil}, \citenamefont
		{Reeve}, \citenamefont {Weigand}, \citenamefont {Van~Waeyenberge},
		\citenamefont {Sch{\"u}tz}, \citenamefont {Everschor-Sitte}, \citenamefont
		{Beach},\ and\ \citenamefont {Kl{\"a}ui}}]{Litzius2020}%
	\BibitemOpen
	\bibfield  {author} {\bibinfo {author} {\bibfnamefont {K.}~\bibnamefont
			{Litzius}}, \bibinfo {author} {\bibfnamefont {J.}~\bibnamefont {Leliaert}},
		\bibinfo {author} {\bibfnamefont {P.}~\bibnamefont {Bassirian}}, \bibinfo
		{author} {\bibfnamefont {D.}~\bibnamefont {Rodrigues}}, \bibinfo {author}
		{\bibfnamefont {S.}~\bibnamefont {Kromin}}, \bibinfo {author} {\bibfnamefont
			{I.}~\bibnamefont {Lemesh}}, \bibinfo {author} {\bibfnamefont
			{J.}~\bibnamefont {Zazvorka}}, \bibinfo {author} {\bibfnamefont {K.-J.}\
			\bibnamefont {Lee}}, \bibinfo {author} {\bibfnamefont {J.}~\bibnamefont
			{Mulkers}}, \bibinfo {author} {\bibfnamefont {N.}~\bibnamefont {Kerber}},
		\bibinfo {author} {\bibfnamefont {D.}~\bibnamefont {Heinze}}, \bibinfo
		{author} {\bibfnamefont {N.}~\bibnamefont {Keil}}, \bibinfo {author}
		{\bibfnamefont {R.~M.}\ \bibnamefont {Reeve}}, \bibinfo {author}
		{\bibfnamefont {M.}~\bibnamefont {Weigand}}, \bibinfo {author} {\bibfnamefont
			{B.}~\bibnamefont {Van~Waeyenberge}}, \bibinfo {author} {\bibfnamefont
			{G.}~\bibnamefont {Sch{\"u}tz}}, \bibinfo {author} {\bibfnamefont
			{K.}~\bibnamefont {Everschor-Sitte}}, \bibinfo {author} {\bibfnamefont
			{G.~S.~D.}\ \bibnamefont {Beach}}, \ and\ \bibinfo {author} {\bibfnamefont
			{M.}~\bibnamefont {Kl{\"a}ui}},\ }\bibfield  {title} {\enquote {\bibinfo
			{title} {The role of temperature and drive current in skyrmion dynamics},}\
	}\href {\doibase 10.1038/s41928-019-0359-2} {\bibfield  {journal} {\bibinfo
			{journal} {Nature Electronics}\ }\textbf {\bibinfo {volume} {3}},\ \bibinfo
		{pages} {30--36} (\bibinfo {year} {2020})}\BibitemShut {NoStop}%
	\bibitem [{\citenamefont {He}\ and\ \citenamefont {Fan}(2017)}]{He2017}%
	\BibitemOpen
	\bibfield  {author} {\bibinfo {author} {\bibfnamefont {Z.}~\bibnamefont
			{He}}\ and\ \bibinfo {author} {\bibfnamefont {D.}~\bibnamefont {Fan}},\
	}\bibfield  {title} {\enquote {\bibinfo {title} {A tunable magnetic skyrmion
				neuron cluster for energy efficient artificial neural network},}\ }in\ \href
	{\doibase 10.23919/DATE.2017.7927015} {\emph {\bibinfo {booktitle} {Design,
				Automation \& Test in Europe Conference \& Exhibition (DATE), 2017}}}\
	(\bibinfo {year} {2017})\ pp.\ \bibinfo {pages} {350--355}\BibitemShut
	{NoStop}%
	\bibitem [{\citenamefont {Raymenants}\ \emph {et~al.}(2021)\citenamefont
		{Raymenants}, \citenamefont {Bultynck}, \citenamefont {Wan}, \citenamefont
		{Devolder}, \citenamefont {Garello}, \citenamefont {Souriau}, \citenamefont
		{Thiam}, \citenamefont {Tsvetanova}, \citenamefont {Canvel}, \citenamefont
		{Nikonov}, \citenamefont {Young}, \citenamefont {Heyns}, \citenamefont
		{Soree}, \citenamefont {Asselberghs}, \citenamefont {Radu}, \citenamefont
		{Couet},\ and\ \citenamefont {Nguyen}}]{Raymenants2021}%
	\BibitemOpen
	\bibfield  {author} {\bibinfo {author} {\bibfnamefont {E.}~\bibnamefont
			{Raymenants}}, \bibinfo {author} {\bibfnamefont {O.}~\bibnamefont
			{Bultynck}}, \bibinfo {author} {\bibfnamefont {D.}~\bibnamefont {Wan}},
		\bibinfo {author} {\bibfnamefont {T.}~\bibnamefont {Devolder}}, \bibinfo
		{author} {\bibfnamefont {K.}~\bibnamefont {Garello}}, \bibinfo {author}
		{\bibfnamefont {L.}~\bibnamefont {Souriau}}, \bibinfo {author} {\bibfnamefont
			{A.}~\bibnamefont {Thiam}}, \bibinfo {author} {\bibfnamefont
			{D.}~\bibnamefont {Tsvetanova}}, \bibinfo {author} {\bibfnamefont
			{Y.}~\bibnamefont {Canvel}}, \bibinfo {author} {\bibfnamefont {D.~E.}\
			\bibnamefont {Nikonov}}, \bibinfo {author} {\bibfnamefont {I.~A.}\
			\bibnamefont {Young}}, \bibinfo {author} {\bibfnamefont {M.}~\bibnamefont
			{Heyns}}, \bibinfo {author} {\bibfnamefont {B.}~\bibnamefont {Soree}},
		\bibinfo {author} {\bibfnamefont {I.}~\bibnamefont {Asselberghs}}, \bibinfo
		{author} {\bibfnamefont {I.}~\bibnamefont {Radu}}, \bibinfo {author}
		{\bibfnamefont {S.}~\bibnamefont {Couet}}, \ and\ \bibinfo {author}
		{\bibfnamefont {V.~D.}\ \bibnamefont {Nguyen}},\ }\bibfield  {title}
	{\enquote {\bibinfo {title} {Nanoscale domain wall devices with magnetic
				tunnel junction read and write},}\ }\href {\doibase
		10.1038/s41928-021-00593-x} {\bibfield  {journal} {\bibinfo  {journal}
			{Nature Electronics}\ }\textbf {\bibinfo {volume} {4}},\ \bibinfo {pages}
		{392--398} (\bibinfo {year} {2021})}\BibitemShut {NoStop}%
\end{thebibliography}
\end{document}